\documentclass[manuscript]{aastex631}
\usepackage[T1]{fontenc}
\usepackage[utf8]{inputenc}
\usepackage{varwidth}
\usepackage{amsbsy}
\usepackage{graphicx}
\usepackage{esint}

\makeatletter

\usepackage{varwidth}
\usepackage{amsbsy}

\submitjournal{ApJ}

\shorttitle{}
\shortauthors{}

\makeatother

\begin{document}

\title{Helicity Fluxes and Hemispheric Helicity Rule 
of Active Regions Emerging from the Convection Zone Dynamo}

\author[0000-0001-9884-1147]{Valery V. Pipin}
\affiliation{National Astronomical Observatories, Chinese Academy of Sciences,
20A Datun Road, Chaoyang District, Beijing, China}
\affiliation{Institute of Solar-Terrestrial Physics, Russian Academy of Sciences,
Irkutsk, 664033, Russia}

\author[0000-0002-2967-4522]{Shangbin Yang}

\affiliation{National Astronomical Observatories, Chinese Academy of Sciences,
20A Datun Road, Chaoyang District, Beijing, China}

\author[0000-0003-0364-4883]{Alexander G. Kosovichev}

\affiliation{New Jersey Institute of Technology, Newark, NJ 07102, USA}
\begin{abstract}
Using a 3D non-linear mean-field solar dynamo model, we investigate
the magnetic helicity flux and magnetic twist, and tilt parameters
of bipolar magnetic regions (BMRs) emerging from the solar convection
zone due to the magnetic buoyancy instability. The twist and tilt of  the BMR magnetic field are modeled as a result of an effective electromotive
force along the rising part of the toroidal magnetic field. This force
generates the poloidal field that tilts the whole magnetic configuration.
We find that variations of BMR's twist and tilt determine the magnitude
and the sign of the magnetic helicity flux on the solar surface. The model
shows that the helicity flux associated with the BMR's tilt/twist
is the dominant contribution to the BMR helicity at the beginning
of the BMR's evolution, while the effect of differential rotation
is the main source of the helicity flux at the final stage of the
BMR's evolution. We discuss the implications of these effects on the
basic properties and variations of the hemispheric helicity rule of
active regions on the solar surface. 
\end{abstract}


\section{Introduction}

\label{sec:Introduction} Magnetic helicity balance plays an important
role in the dynamo processes in the solar and stellar convection zones.
In particular, the nonlinear saturation of the turbulent dynamo significantly
depends on the evolution of the magnetic helicity and its expulsion
from the dynamo domain \citep{Kleeorin1982,BRetal23}. Moreover, the magnetic helicity
flux from the depth of the convection zone can affect the activity
phenomena in the chromosphere and corona. For example, the amount
of helicity stored in solar active regions affects their flare and
coronal mass ejection (CME) productivity \citep[e.g.,][]{Berger2000,PariatAnDev2009ApJ,Georgoulis2009,Toriumi2019}.

It was found that the solar differential rotation provides a major
effect on the rate of helicity production, both in the flaring and
CME activity of individual active regions and in the progression of
solar activity cycles \citep{Berger2000,Hawkes2019AA}. This supports
the basic dynamo scenario of \citet{Parker1955}, suggesting that
the differential rotation and turbulent generation of large-scale
magnetic fields are the main sources of the magnetic energy generated
in the solar dynamo. This model predicted that the dynamo-generated
magnetic field migrates in radius and latitude in the form of dynamo
waves.
It showed that magnetic stresses and modulation of the turbulent heat flux,
associated with these waves, {result in 11-year variations of the differential rotation
(“torsional oscillations”) which are characterized by an extended 22-year mode that propagates during two solar cycles from the polar regions to the equator \citep{Kosovichev2019,PK19,Manda2024}. Similarly, the model reproduces variations of the meridional circulation, also in agreement with helioseismology results \citep[e.g.,][]{Komm2018,Pipin2020,Getling2021,Getling2025}.}

The emergence of the tilted bipolar magnetic regions (BMR), together
with the effects of the cyclonic convection motions (associated with
the so-called $\alpha$-effect), represents the turbulent dynamo generation
of the large-scale magnetic field observed on the surface of the Sun.
{Both effects are related to the production of the helical magnetic
field and therefore affect the magnetic helicity fluxes from the solar
convection zone. }Previous studies (e.g., \citealp{Kleeorin2000,Blackman2003,Brandenburg2005b})
suggested an important role for turbulent magnetic helicity fluxes
for the large-scale dynamo.

Another important aspect of the problem is the hemispheric helicity
rule (hereafter HHR). It states that the electric current helicity,
which is a proxy of magnetic helicity, is predominantly negative in
the Northern hemisphere and positive in the Southern hemisphere. In
other words, the magnetic field of bipolar magnetic regions (BMRs) 
is twisted counter-clockwise
in the Northern hemisphere and clockwise in the Southern hemisphere.
Starting from the results of \citet{Seehafer1990}, \citet{Pevtsov1994}
and \citet{Bao1998}, the hemispheric helicity rule is a well-established
statistical pattern of the solar active regions. However, there are
significant fluctuations \citep{Zhang2024GMS}. In particular, observations
show that solar active regions can violate the hemispheric helicity
rule mostly during the initial phase of active region emergence \citep{Kutsenko2019MNRAS}.
The mean-field dynamo models attempt to relate the HHR with the sign of the $\alpha$-effect and magnetic helicity conservation
\citep{Sokoloff2006,Pipin2013c}. The surface flux-transport models
interpret observations of the HHR in a different way. For example,
\citet{Prior2014ApJ} modeled HHR as an effect of the differential
rotation acting on the initial random distribution of the helical
BMRs. The reader can find more information on solar magnetic helicity
and beyond in reviews published in \citet{Kuzanyan2024GMS}. Using
a surface flux-transport model, \citet{Hawkes2019AA} found that the
magnitude of the helicity flux due to the decay of active regions
is about two orders of magnitude lower than that of the helicity flux
produced by the differential rotation. However, the flux-transport
models do not take into account the radial dependence of the magnetic
field distributions and the BMR emergence in the convection zone (\citealp{Yeates2023a,Pipin2024SoPh}).
This can lead to an underestimation of the BMR role in the magnetic
helicity budget and consequently the helicity flux from the photosphere to the solar corona.
Nevertheless, it is important that the magnetic helicity flux, initiated by the BMR's emergence and evolution, and the HHR  can be closely related.

Our general goal is to evaluate the contribution of BMRs to the total
magnetic helicity balance. For this purpose, we use the non-linear
3D MHD dynamo model of \citet{PKT23}, which addresses the emergence
and evolution of BMR simultaneously with the global dynamo in the
solar convection zone. We calculate the helicity flux initiated
by BMRs and also the latitudinal distribution of the magnetic twist parameters
of BMRs. The dynamo model allows us to estimate directly the
contributions of the helicity production rate on the solar
surface caused by the large-scale flows and the evolution of the bipolar
active regions. Our plan is as follows. Section 2 discusses some aspects
of the dynamo model and the evolution equation for the helicity rate.
In Section 3, we calculate the surface helicity flux using typical
configurations for emerging BMRs. Then, we calculate the HHR for the
BMR's twist parameters and helicity flux. The paper ends with
a discussion and conclusions.

\section{Magnetic Helicity Balance}

The total magnetic helicity inside the convection zone can be defined
via the volume integral,
\begin{equation}
\mathcal{H}_{V}=\int\boldsymbol{A\cdot B}\mathrm{dV,}\label{eq:Hv}
\end{equation}
 where $\boldsymbol{A}$ is the magnetic vector potential, $\boldsymbol{B}=\nabla\times\boldsymbol{A}$. Hereafter, we assume the volume integral is calculated over the bulk of the convection zone.
{ It is worth noting that the helicity integral of  the large-scale dynamo-generated magnetic fields turns to zero when integrating over the hemispheres of the Sun. In this paper, we utilize the integral form of the helicity conservation law to derive the evolution equation for the helicity of the small-scale magnetic field, also, see \cite{Hubbard2012}.}
Generally, the vector-potential is defined only up to a gauge transformation,
$\boldsymbol{A}\rightarrow\boldsymbol{A}+\boldsymbol{\nabla}g$, where
$g$ is an arbitrary scalar. For the large-scale dynamo models, the
uncertainty is cured by the decomposition of the magnetic field into a sum
of the toroidal, $\boldsymbol{B}_{T}$, and poloidal components, $\boldsymbol{B}_{P}$,
\citep{Krause1980}, which are decomposed further following \citet{CandrKend1957ApJ}:
\begin{eqnarray}
\boldsymbol{B} & = & \boldsymbol{B}_{T}+\boldsymbol{B}_{P}=\boldsymbol{\nabla}\times\boldsymbol{r}T\left(\boldsymbol{r},t\right)+\boldsymbol{\nabla}\times\boldsymbol{\boldsymbol{\nabla}\times r}S\left(\boldsymbol{r},t\right)\label{ChKe}
\end{eqnarray}
{where the first term in the RHS corresponds to $\boldsymbol{B}_{T}$,
and $T$ and $S$ are scalars, which are called superpotentials.
The superpotentials also have gauge uncertainty. However,
in this case, the arbitrary scalars, which take part in the gauge
transformation, depend on the radial coordinate only. This uncertainty
can be removed if we consider the appropriate integral averaging of
$T$ and $S$. The procedure is particularly simple in the case of the
spherical dynamo models, see more details in Appendix~\ref{A} and
the book by \citet{Krause1980}.  The helicity integral, $\mathcal{H}_{V}$, measures the linkage of, $\boldsymbol{B}_{T}$ and $\boldsymbol{B}_{P}$
in the volume \citep{Berger2018}.}

{We employ the Electromagnetic units system throughout the paper, and following the Faraday law, }
\begin{equation}
\frac{\partial\boldsymbol{B}}{\partial t}=-\nabla\times\boldsymbol{E},\label{eq:maxE}
\end{equation}
where {$\boldsymbol{E}$} is the electric field, determine the helicity rate in the bulk of the convection zone: 
\begin{eqnarray}
\frac{\mathrm{d}\mathcal{H}_{V}}{\mathrm{d}t} & = & \int\left(\frac{\partial\boldsymbol{A}}{\partial t}\cdot\boldsymbol{B}+\boldsymbol{A}\cdot\frac{\partial\boldsymbol{B}}{\partial t}\right)\mathrm{dV}=\int\left(2\boldsymbol{A}\cdot\frac{\partial\boldsymbol{B}}{\partial t}+\nabla\cdot\left(\boldsymbol{A}\times\frac{\partial\boldsymbol{A}}{\partial t}\right)\right)\mathrm{dV}\label{BergField84}\\
 & = & -2\int\boldsymbol{E}\cdot\boldsymbol{B}dV+2\oint d\boldsymbol{S}\cdot\left(\boldsymbol{A}\times\boldsymbol{E}\right)+\oint d\boldsymbol{S}\cdot\left(\boldsymbol{A}\times\frac{\partial\boldsymbol{A}}{\partial t}\right).\label{eq:BFb}
\end{eqnarray}
It is noteworthy that the last integral in this formula is identically
zero \citep{Berger2018}. We keep it because in the dynamo equations,
$\left(\boldsymbol{A}\times{\displaystyle \frac{\partial\boldsymbol{A}}{\partial t}}\right)$
has a counterpart in $\left(\boldsymbol{A}\times\boldsymbol{E}\right)$.
Taking into account Ohm's law, 
\begin{equation}
\boldsymbol{E}=-\boldsymbol{v}\times\boldsymbol{B}+\eta\boldsymbol{J},\label{eq:Ohm}
\end{equation}
where  $\boldsymbol{v}$ is the plasma velocity, $\boldsymbol{J}$ is the electric current
density, and $\eta$ is the microscopic diffusivity, we get the helicity
rate in the volume of the convection zone: 
\begin{eqnarray}
\frac{\mathrm{d}\mathcal{H}_{V}}{\mathrm{d}t} & = & -2\eta\int\boldsymbol{B\cdot}\boldsymbol{J}dV+2\oint d\boldsymbol{S}\cdot\boldsymbol{B}\left(\boldsymbol{A}\cdot\boldsymbol{v}\right)-2\oint d\boldsymbol{S}\cdot\boldsymbol{v}\left(\boldsymbol{A}\cdot\boldsymbol{B}\right)\label{eq:HR0}\\
 &  & +2\eta\oint d\boldsymbol{S}\cdot\left(\boldsymbol{A}\times\boldsymbol{J}\right)+\oint d\boldsymbol{S}\cdot\left(\boldsymbol{A}\times\frac{\partial\boldsymbol{A}}{\partial t}\right).
\end{eqnarray}
Next, following the standard approach of the mean-field magnetohydrodynamics,
we decompose the induction vector of the magnetic field, $\boldsymbol{B}$,
and its vector-potential $\boldsymbol{A}$, into the mean and fluctuating
parts, 
\begin{eqnarray}
\boldsymbol{B} & = & \left\langle \boldsymbol{B}\right\rangle +\boldsymbol{b},\label{eq:dec}\\
\boldsymbol{A} & = & \left\langle \boldsymbol{A}\right\rangle +\boldsymbol{a},\nonumber \\
\boldsymbol{v} & = & \left\langle \boldsymbol{U}\right\rangle +\boldsymbol{u},\nonumber \\
\boldsymbol{J} & = & \left\langle \boldsymbol{J}\right\rangle +\boldsymbol{j}\nonumber 
\end{eqnarray}
where the small letters denote the turbulent fluctuations and the
angular brackets denote the averaging over the ensemble of fluctuations.
Substituting these decompositions into Eq.(\ref{eq:BFb})
and averaging over the ensemble of fluctuations, we get, 
\begin{eqnarray}
\frac{\mathrm{d}\mathcal{H}_{V}}{\mathrm{d}t} & = & \frac{\mathrm{d}}{\mathrm{d}t}\int\left(\left\langle \boldsymbol{a\cdot}\boldsymbol{b}\right\rangle +\left\langle \boldsymbol{A}\right\rangle \cdot\left\langle \boldsymbol{B}\right\rangle \right)\mathrm{d}V=-2\eta\int\left(\left\langle \boldsymbol{b}\cdot\boldsymbol{j}\right\rangle +\left(\left\langle \boldsymbol{B}\right\rangle \cdot\left\langle \boldsymbol{J}\right\rangle \right)\right)\mathrm{d}V\label{eq:intcons}\\
 & - & \ointop d\boldsymbol{S}\cdot\boldsymbol{F}^{\left\langle ab\right\rangle }+2\eta\oint d\boldsymbol{S}\cdot\left(\left\langle \boldsymbol{a}\times\boldsymbol{j}\right\rangle +\left\langle \boldsymbol{A}\right\rangle \times\left\langle \boldsymbol{J}\right\rangle \right)+\oint d\boldsymbol{S}\cdot\left(\left\langle \boldsymbol{A}\right\rangle \times\frac{\partial\left\langle \boldsymbol{A}\right\rangle }{\partial t}\right),\nonumber 
\end{eqnarray}
where $\boldsymbol{F}^{\left\langle ab\right\rangle }$ is the  flux of the turbulent magnetic helicity density, $\left\langle \boldsymbol{a\cdot}\boldsymbol{b}\right\rangle$.
{This equation shows that the total helicity rate in the volume is
only due to the Ohmic dissipation of the current helicity, and the
turbulent helicity flux $\boldsymbol{F}^{\left\langle ab\right\rangle}$ 
through the dynamo domain boundaries.}
In the mean-field theory, the general expression of $\boldsymbol{F}^{\left\langle ab\right\rangle }$
is complicated (see, \citealt{Kleeorin2022,Subramanian2023ApJ}).
It includes the products of the large-scale flow, $\left\langle \boldsymbol{U}\right\rangle $
, magnetic field, $\left\langle \boldsymbol{B}\right\rangle $ and
its vector potential $\left\langle \boldsymbol{A}\right\rangle $
with the second moments of the turbulent fields, and the triple-order
moments of the turbulent fields. In our study, we approximate it by the effect of turbulent diffusion, 
\begin{equation}
\boldsymbol{F}^{\left\langle ab\right\rangle }=-\eta_{\chi}\boldsymbol{\nabla}\left\langle \boldsymbol{a\cdot}\boldsymbol{b}\right\rangle .\label{eq:hdf}
\end{equation}
Following the results of \citet{Mitra2010,Kleeorin2022}, we set $\eta_{\chi}=\frac{1}{10}\eta_{T}$,
where $\eta_{T}$ is the amplitude of the magnetic eddy diffusivity.
The latter is determined with the help of the analytical results of
the mean-field theory and the standard
mixing-length approximation for the convective zone turbulent flows.
{Except for $\boldsymbol{F}^{\left\langle ab\right\rangle }$, the
second line of Eq.(\ref{eq:intcons}) contains the Fickian-type
fluxes of the small-scale and large-scale helicity due to the Ohmic diffusion, i.e., the contributions
like, $\eta\boldsymbol{\nabla}\left\langle \boldsymbol{a\cdot}\boldsymbol{b}\right\rangle $
and $\eta\boldsymbol{\nabla}\left\langle \boldsymbol{A}\right\rangle \cdot\left\langle \boldsymbol{B}\right\rangle $
(cf., Eq.(\ref{fluxEta}) and Sec.~\ref{subsec:Model}). However, these
contributions are much smaller in comparison to the turbulent diffusion, and, therefore, we neglect them in our analysis.}

{The evolution equation for the small-scale helicity can be obtained from
Eq(\ref{eq:intcons}) using the mean-field induction equation,
\begin{equation}
\frac{\partial\left\langle \boldsymbol{B}\right\rangle }{\partial t}=\nabla\times\left(\boldsymbol{\mathcal{E}}+\left\langle \boldsymbol{U}\right\rangle \times\left\langle \boldsymbol{B}\right\rangle \right),\label{eq:MFE}
\end{equation}
where $\boldsymbol{\mathcal{E}}=\left\langle \boldsymbol{u}\times\boldsymbol{b}\right\rangle$
is the mean electromotive force of the turbulent flows. We describe the mathematical details in Appendix \ref{A}. The final result is as follows,
\begin{eqnarray}
\frac{\mathrm{d}}{\mathrm{d}t}\int\left\langle \boldsymbol{a\cdot}\boldsymbol{b}\right\rangle dV & = & -2\int\left(\boldsymbol{\mathcal{E}}\cdot\left\langle \boldsymbol{B}\right\rangle \right)dV-\int\frac{\left\langle \boldsymbol{a\cdot}\boldsymbol{b}\right\rangle }{R_{m}\tau_{c}}dV+2\eta\oint d\boldsymbol{S}\cdot\left\langle \boldsymbol{a}\times\boldsymbol{j}\right\rangle \label{eq:ab-2-1}\label{eq:ab2}\\
 & - & \oint d\boldsymbol{S}\cdot\boldsymbol{F}^{\left\langle ab\right\rangle }+\oint d\boldsymbol{S}\cdot\left\langle \boldsymbol{U}\right\rangle \left(\left\langle \boldsymbol{A}\right\rangle \cdot\left\langle \boldsymbol{B}\right\rangle \right)\nonumber\\
 & - & 2\oint d\boldsymbol{S}\cdot\left(\boldsymbol{\mathcal{E}}\times\left\langle \boldsymbol{A}\right\rangle \right)-2\oint d\boldsymbol{S}\cdot\left\langle \boldsymbol{B}\right\rangle \left(\left\langle \boldsymbol{A}\right\rangle \cdot\left\langle \boldsymbol{U}\right\rangle \right),\nonumber
\end{eqnarray}
where, $R_{m}$ is the turbulent magnetic Reynolds number, $\tau_{c}$
is the typical convective turnover time of the turbulent flows. Here,
we employ the result of \citet{Kleeorin1999} for isotropic turbulence,
$2\eta\left\langle \boldsymbol{b}\cdot\boldsymbol{j}\right\rangle ={\displaystyle \frac{\left\langle \boldsymbol{a\cdot}\boldsymbol{b}\right\rangle }{R_{m}\tau_{c}}}$.
In this study, we assume that the normal to the surface component of
the large-scale flow, $\left\langle \boldsymbol{U}\right\rangle $,
is zero at the top boundary. { In the dynamo model, we use the differential form of Eq.(\ref{eq:ab2}). The specific area of the surface integrals is described in Section 4.}

The term -2$\oint d\boldsymbol{S}\cdot\left(\boldsymbol{\mathcal{E}}\times\left\langle \boldsymbol{A}\right\rangle \right)$
represents the helicity flux initiated by the turbulent processes
in the large-scale dynamo. \citet{Pipin2013c} found that this helicity flux results in the small-scale magnetic helicity density evolution following the large-scale dynamo wave.
This alleviates the non-linear saturation (catastrophic quenching) of the $\alpha$-effect. \cite{DSordo2013MN} and \cite{Brandenburg2018AN} investigated this flux using direct numerical simulations and  found that it was difficult to confirm this effect due to the limited numerical resolution. }
The term $-2\oint d\boldsymbol{S}\cdot\left\langle \boldsymbol{B}\right\rangle \left(\left\langle \boldsymbol{A}\right\rangle \cdot\left\langle \boldsymbol{U}\right\rangle \right)$
stands for effects of the large-scale flow, i.e., the differential
rotation and meridional circulation \citep{Berger2000,Hawkes2019AA}.
Our goal is to study the contribution of the bipolar active regions
to these helicity fluxes. 

To achieve this goal, we consider the dynamo model with emerging active regions proposed by \citet{PKT23}.
In this model, the evolution equation for the mean magnetic induction
vector, $\left\langle \boldsymbol{B}\right\rangle $, describes both
the dynamo-generated large-scale magnetic field and the magnetic field
of active regions that are formed from the large-scale toroidal
magnetic field due to the magnetic buoyancy instability. Such formulation of the mean-field dynamo model is
possible by considering mean nonaxisymmetric magnetic fields.
In the model, we approximate the magnetic configurations of active
regions in the form of bipolar magnetic structures. Observations show
that the contribution of bipolar-like active regions to the total
unsigned flux of the photospheric radial magnetic fields is less than
10 percent \citep{Nagovitsyn2016a,Pevtsov2021}. Moreover, the flux
distribution in the solar active regions shows a rich diversity of
the magnetic patterns \citep{Abramenko2023m}. The above arguments
show the limitations and the main source of uncertainty in the comparison
of the model with observations. 

We consider the mean magnetic field induction equation (Eq.~\ref{eq:MFE}) for the highly
conductive media with the addition of the effects of the bipolar magnetic
regions.  In this equation, the electromotive force $\mathcal{E}$ contains both the mean-field turbulent
effects and the generation terms for BMRs, defined in Appendix \ref{A}
and the next section. The mean large-scale flow velocity, $\left\langle \boldsymbol{U}\right\rangle $,
represents the differential rotation and the meridional circulation.
It is calculated consistently by solving Eq.\,(\ref{eq:MFE}) together
with the equations that describe the angular momentum balance, the
meridional circulation, the mean-field heat transport, and the integral
balance of the magnetic helicity in the bulk of the solar convection
zone \citep{PK24}. We use the harmonic field approximation \citep{Bonanno2016}
outside the dynamo domain, which is more suitable for modeling the
magnetic helicity because, unlike the usual potential field approximation,
it does not suppress the contributions from the tilt and twist of
BMRs on the surface. In this case, we employ the standard boundary
conditions: continuity of the normal component of the magnetic field
and the tangential component of the mean electromotive force (see
Appendix B).

\begin{figure}
\centering \includegraphics[width=0.95\textwidth]{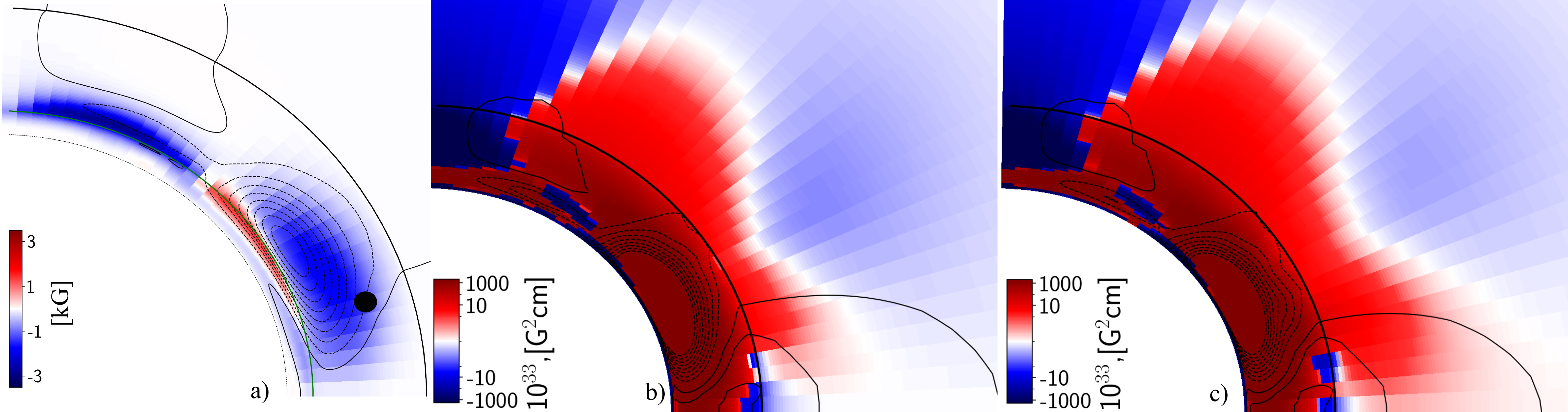}

\caption{{A snapshot of the large-scale axisymmetric magnetic field and magnetic helicity density in the
northern hemisphere of the Sun: a) the color image shows the toroidal
magnetic field, and the contour lines of the axisymmetric vector potential
show the poloidal magnetic field lines; the black circle shows the
position of the BMR initiation; b) the color image show the magnetic helicity density of axisymmetric magnetic field at the same time as in panel (a), and the contour lines are the same as in panel (a)); c) the same as b) at the end of the run E6, which included the initial axisymmetric poloidal magnetic field, see Table 1.}}\label{fig1}
\end{figure}

\section{Model of tilted/twisted bipolar magnetic regions}

\label{subsec:Model}

The mean electromotive force, $\boldsymbol{\mathcal{E}}=\left\langle \boldsymbol{u}\times\boldsymbol{b}\right\rangle $,
represents the effects of turbulent flows and magnetic fields on the large-scale magnetic field induction. We formulate it as follows, 
\begin{equation}
\mathcal{E}_{i}=\left(\alpha_{ij}+\gamma_{ij}\right)\left\langle B\right\rangle _{j}-\eta_{ijk}\nabla_{j}\left\langle B\right\rangle _{k}+{\mathcal{E}}_i^{\mathrm{(BMR)}},\label{eq:emf-1}
\end{equation}
 where   $\alpha_{ij}$ describes the turbulent generation by the
hydrodynamic magnetic helicity (the global dynamo $\alpha$-effect), $\gamma_{ij}$ is the turbulent
pumping, $\eta_{ijk}$ is the eddy magnetic diffusivity tensor, and $\boldsymbol{\mathcal{E}}^{(\mathrm{BMR})}$ models the emergence of the tilted/twisted bipolar active regions, see details
in Appendix A. The additional term of the mean electromotive force,
$\boldsymbol{\mathcal{E}}^{(\mathrm{BMR})}$, is formulated as follows
\citep{PKT23}: 
\begin{equation}
\boldsymbol{\mathcal{E}}^{(\mathrm{BMR})}=\alpha_{\beta}^{\rm BMR}\left\langle \boldsymbol{B}\right\rangle +V_{\beta}\left(\hat{\boldsymbol{r}}\times\left\langle \mathbf{B}\right\rangle \right),\label{eq:ebmrt}
\end{equation}
where the first term takes into account the BMR's tilt/twist and the
second term models the rise of the magnetic region to the surface
in the bipolar form with velocity $V_{\beta}$. In our basic scenario,
the term $\alpha_{\beta}^{\rm BMR}\left\langle B_{\phi}\right\rangle $ induces
an effective electromotive force along the rising part of the toroidal
magnetic field. This electromotive force generates the poloidal magnetic
field, which tilts the whole magnetic configuration of BMR. If we
leave the toroidal magnetic field at rest (no rise), then this additional
small-scale poloidal magnetic field results in the twisted magnetic
field configuration. Therefore, it makes sense to divide the BMR
formation process, described by Eq.\,(\ref{eq:ebmrt}), into the
two corresponding parts: 
\begin{eqnarray}
\boldsymbol{\mathcal{E}}^{(\mathrm{BMR})} & = & \boldsymbol{\mathcal{E}}_{\mathrm{1}}+\boldsymbol{\mathcal{E}}_{2},\label{eq:twtlt}\\
\boldsymbol{\mathcal{E}}_{\mathrm{1}} & = & \alpha_{\beta}^{\rm BMR}\left\langle \boldsymbol{B}\right\rangle \xi_{1}(t,\boldsymbol{r})\nonumber \\
\boldsymbol{\mathcal{E}}_{\mathrm{2}} & = & V_{\beta}\left(\hat{\boldsymbol{r}}\times\left\langle \mathbf{B}\right\rangle \right)\xi_{2}(t,\boldsymbol{r})\nonumber 
\end{eqnarray}
where functions $\xi_{1}$ and $\xi_{2}$ describe the spatio-temporal
parameters of the initial perturbations of the magnetic buoyancy instability.
The magnetic buoyancy velocity, $V_{\beta}$, includes the turbulent
and mean-field buoyancy effects \citep{Kitchatinov1993}: 
\begin{eqnarray}
V_{\beta} & = & \frac{\alpha_{\mathrm{MLT}}u_{c}}{\gamma}\beta^{2}\mathcal{H}\left(\beta\right)\label{eq:bu}
\end{eqnarray}
where function $\mathcal{H}\left(\beta\right)$ describes the quenching
effect of the magnetic tension (see the above-cited paper). Also,
$\alpha_{\mathrm{MLT}}=1.9$ is the mixing length theory parameter,
$u_{c}$ is the RMS convective velocity, and $\gamma$ is the adiabatic
constant. All these parameters are taken from the results of the standard
MESA model for the Sun \citep{Paxton2011}. Following \citet{P22},
we define 
\begin{equation}
\alpha_{\beta}^{\rm BMR}=C_{\alpha\beta}\cos\theta V_{\beta}\psi_{\alpha}(\beta).\label{eq:ab-1}
\end{equation}
Here, the parameter $C_{\alpha\beta}$ determines the magnitude of
tilt/twist of the BMR for a given latitude. The function, $\psi_{\alpha}\left(\beta\right)$,
where $\mathrm{\beta=\left|\left\langle \mathbf{B}\right\rangle \right|/\sqrt{4\pi\overline{\rho}u_{c}^{2}}}$,
describes the algebraic quenching of the $\alpha$ effect. We define
the functions $\xi_{1,2}(t,\mathbf{r})$ in the same way as \citet{PKT23}.
Their description is given in Appendix C. {A similar source of
the $\alpha$-effect was suggested earlier by \citet{Ferriz1994}. Our form of the $\alpha$-effect leverages their idea by utilizing the explicit relationship between the amplitude of $\alpha_{\beta}$ and the magnetic buoyancy velocity. Moreover, in the dynamo model, the radial and latitudinal positions of the unstable point are calculated using the instability parameter connected with the Parker buoyancy instability (see, \citealp{P22}).} 
In our model, the typical amplitude of $\alpha_{\beta}^{\rm BMR}$ is about 5-10 $m/s$. This is of the same order of magnitude as the global dynamo $\alpha$-affect in the upper
part of the convection zone (see Figure 3 in the above-cited paper). According to \cite{Choudhuri1992} and \cite{Hoyng1993}, such
strong fluctuations can be possible if we take into account the local character of the BMR's formation.

{We did not investigate whether this effect could generate a large-scale dynamo on its own. Clearly, a solar-like BMR can be produced when a sufficiently strong seed toroidal magnetic field is present. Its action on a weak poloidal field is consistent with the standard mean-field $\alpha$ effect. \cite{P22} found that for a given parameter   $C_{\alpha\beta}$, the amplitudes of the poloidal and toroidal magnetic fields produced by the BMRs are larger by about ten percent of their values in the mean-field global axisymmetric dynamo model without BMRs for the same $\alpha$-effect parameter $C_{\alpha\beta}$.
}


Depending on the mutual phase of $\boldsymbol{\mathcal{E}}_{\mathrm{1}}$
and $\boldsymbol{\mathcal{E}}_{2}$, the sign of $C_{\alpha\beta}$
and the employed boundary conditions, we can distinguish several interesting
cases for the study. We assume that the emergence phase, which is
associated with magnetic buoyancy, $\boldsymbol{\mathcal{E}}_{2}$,
can start either after the action of $\boldsymbol{\mathcal{E}}_{1}$
or simultaneously with it. When $\boldsymbol{\mathcal{E}}_{\mathrm{1}}$
precedes $\boldsymbol{\mathcal{E}}_{2}$, it results in the rise of
a twisted bipolar magnetic field structure. The simultaneous action
of $\boldsymbol{\mathcal{E}}_{\mathrm{1}}$ and $\boldsymbol{\mathcal{E}}_{2}$
produces a tilted and twisted BMR. We can vary the sign and phase
of $\boldsymbol{\mathcal{E}}_{\mathrm{1}}$ to generate the different
signs of twist and tilt. We list the cases in Table 1.

For the source of the BMR initiation, we considered the toroidal magnetic
field in the upper part of the convection zone at the growing stage
of the dynamo cycle, where the condition for the magnetic buoyancy instability is satisfied \citep{PKT23}. Snapshots of the large-scale magnetic field and its helicity density distributions before and after the BMR's emergence are shown in Figure \ref{fig1}, as an example.
{In the model, the large-scale magnetic field is almost antisymmetric
about the equator. This is because the dynamo model employs the mean-field alpha effect, which is slightly above the dynamo instability threshold when the quadrupolar modes are still subcritical. The nonlinear dynamo processes, as well as spontaneous
BMR formation can break the parity of the magnetic field during the
solar cycle. This can also affect the helicity fluxes. Here, we
ignore these effects and consider the magnetic helicity parameters
for a particular stage of the solar cycle with an almost antisymmetric
configuration of the large-scale magnetic field.
} 
{The latitude of the initial perturbation shown by the black circle is fixed at $20^{\circ}$. Inside the convection zone, the helicity density shows opposite signs at low and high latitudes. These signs originate from the dynamo region. We made one of the runs using setup E6 (Table 1) and taking into account the full axisymmetric magnetic field.  Figures \ref{fig1}(b and c) show snapshots of the axisymmetric magnetic field helicity density at the beginning and at the end of the run. This run (E6) employs the harmonic boundary conditions. In the northern hemisphere, we see the injection of the positive helicity at the end of the run. Also, the helicity density in the low corona changes in the near-equatorial regions.  Similarly to the results of \cite{Warnecke2011,Bonanno2016, Bourdin2018ApJ}, the magnetic helicity density of the axisymmetric magnetic field shows an inversion of sign at radius $r\approx 1.7R_\odot$. This effect was found in the analysis of solar wind observations by \citet{Brandenburg2011}. We leave a detailed study of this problem for another paper.} 
{To exclude the effects of interaction of the magnetic field of BMR with the axisymmetric poloidal magnetic field that may exist before the BMR's emergence, we set the initial axisymmetric poloidal magnetic field strength. For models E5 and E6, we make additional runs varying the initiation latitude of the BMR initiation in the range $\pm40^{\circ}$.} 
\begin{figure}
\centering \includegraphics[width=0.95\textwidth]{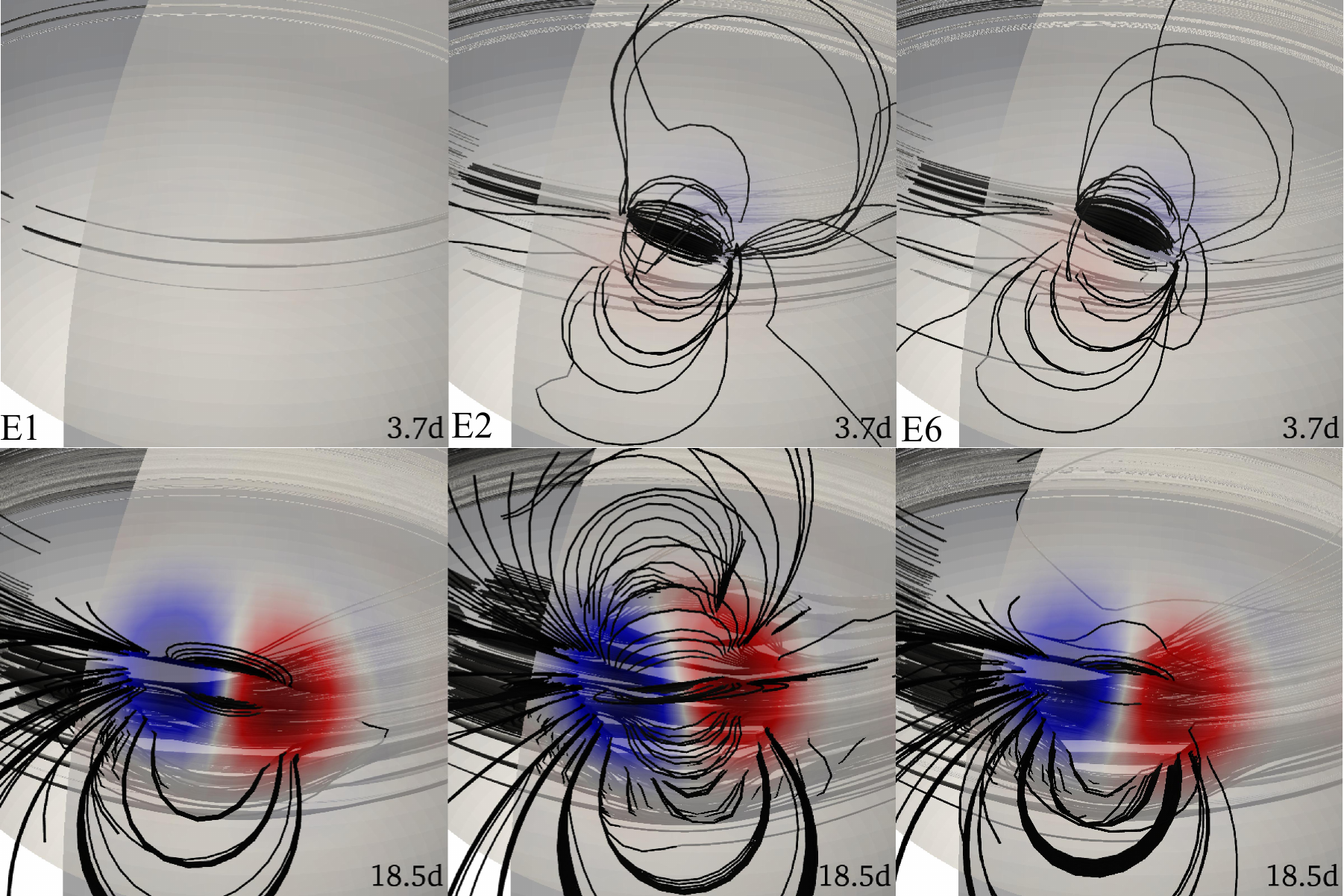}

\caption{The top row shows the snapshots of the magnetic field distribution in
a bipolar magnetic region (BMR) at the beginning of its emergence
at the surface. The bottom row shows the same for the final stage
of the simulation. The columns marked E1, E2, and E6 correspond to
the cases listed in Table 1. The color image shows the radial magnetic
field. {The online version contains an animation of this Figure. The animation illustrates the magnetic field evolution of BMRs, spanning 2 to 19 days, during the evolution of the active regions.}}\label{fig2}
 
\end{figure}

For the simultaneous action of $\boldsymbol{\mathcal{E}}_{\mathrm{1}}$
and $\boldsymbol{\mathcal{E}}_{\mathrm{2}}$, in the case E1, we obtain
a tilted BMR illustrated in Figure \ref{fig2}. The tilt of the BMR
does not change much during emergence in this case. The figure illustrates
two other situations. Case E2 employs the boundary conditions of the
harmonic magnetic field (Appendix B). In this case, $\boldsymbol{\mathcal{E}}_{\mathrm{I}}$
acts during the pre-emergence stage, then it is turned off, and the
BMR starts to rise due to the action of $\boldsymbol{\mathcal{E}}_{\mathrm{2}}$.
This model run shows the clockwise rotation of 180$^{\circ}$ of BMR
during the emerging phase. Case E6 is the same as E1, but with the
harmonic magnetic field boundary conditions. In run E6, the magnetic
polarities rotate anticlockwise by approximately 135$^{\circ}$. These
effects qualitatively correspond to observational results \citep[e.g.,][]{Tian2005SoPh,Sturrock2015AA,Grigoryev2025}.
In both cases, E2 and E6, we find that the polarity pattern is slightly
more elongated along the polarity inversion line than in cases E1
and E4.
\begin{deluxetable*}{ccccccc}\label{tab}
    \tablenum{1}
    \tablecaption{Parameters of the runs.}
    \tablewidth{0pt}
    \tablehead{
        \colhead{Case} & \colhead{Type} & \colhead{BC} & \colhead{$\boldsymbol{\mathcal{E}}_{\mathrm{1}}$} 
        & \colhead{$\boldsymbol{\mathcal{E}}_{\mathrm{2}}$} 
        & \colhead{$C_{\alpha\beta}$} & \colhead{$C_{\alpha\beta}$} \\ 
        \colhead{} & \colhead{BMR} &\colhead{} &\colhead{($\alpha$-effect)} & \colhead{(buoyancy)} &\colhead{'tilt'} &\colhead{ 'twist'}    }
    \startdata
    E1 & tilted & potential & $0<t<\delta t$ & $0<t<\delta t$ & 1 & 0 \\
    E2 & twisted & harmonic & $0<t<\delta t/3$ & $\delta t/3<t<\delta t$ & 0 & 1\\
    E3 & twisted-tilted & potential& $0<t<\delta t/3$ & $\delta t/3<t<\delta t$ & 1 & 1\\
    E4 & tilted & harmonic & $0<t<\delta t$ & $0<t<\delta t$ & -1 & 0\\
    E5 & tilted & harmonic & $0<t<\delta t$ & $0<t<\delta t$ & 1 & 0\\
    E6 & twisted-tilted & harmonic & $0<t<\delta t$ & $0<t<\delta t$ & 1 & 1\\
    \enddata
    \tablecomments{ We set the initial axisymmetric poloidal magnetic field strength to zero. For the runs E5 and E6 we vary the range the BMR's initiation latitude from -40$^{\circ}$ to 40$^{\circ}$. For the run E6 we made the additional run with initiation at 20$^{\circ}$ latitude and the initial axisymmetric poloidal magnetic field geometry as shown in Figure \ref{fig1}.}
\end{deluxetable*}

Using our analytical expressions for the mean electromotive force
(Appendix \ref{A}), we introduce the following definition for the
helicity density flux outward of the dynamo region: 
\begin{eqnarray}
F_{\Omega} & = & 2\left\langle {B}\right\rangle _{r}\left\langle {A}\right\rangle _{\phi}\left\langle {U}\right\rangle _{\phi},\label{eq:EWp}\\
F_{U} & = & 2\left\langle {B}\right\rangle _{r}\left\langle {A}\right\rangle _{\theta}\left\langle {U}\right\rangle _{\theta},\label{eq:EUp}\\
F_{r}^{\left\langle ab\right\rangle } & = & -\eta_{\chi}\left(\hat{\mathbf{r}}\cdot\nabla\right)\left\langle \boldsymbol{a}\cdot\boldsymbol{b}\right\rangle ,\label{eq:fabp}
\end{eqnarray}
where the small-scale helicity density is estimated from Eq.\,(\ref{eq:ab-1}).
To estimate the effect of the BMR's twist, tilt, and the effect
of the turbulent diffusion, we take into account the isotropic structure
of the hydrodynamic $\alpha$-effect and turbulent diffusion near
the solar surface. We define the helicity density flux from the mean
electromotive force as follows, $F_{\mathcal{E}}=F_{\mathcal{\alpha}\beta}+F_{\mathcal{\eta}}$,
where $F_{\mathcal{\alpha}\beta}$ determines the flux from the magnetic
field twist and tilt of the BMR rising from the depth of the convection
zone, 
\begin{equation}
F_{\mathcal{\alpha}\beta}=2\left(\alpha_{\phi\phi}+\alpha_{\beta}^{\rm BMR}\right)\left(\left\langle B\right\rangle _{\phi}\left\langle {A}\right\rangle _{\theta}-\left\langle B\right\rangle _{\theta}\left\langle {A}\right\rangle _{\phi}\right)-2V_{\beta}
\left(\left\langle B\right\rangle _{\theta}\left\langle {A}\right\rangle _{\theta}+\left\langle B\right\rangle _{\phi}\left\langle {A}\right\rangle _{\phi}\right),\label{eq:Fabp}
\end{equation}
where we see that only the horizontal components of the magnetic field
and vector potentials contribute to $F_{\mathcal{\alpha}\beta}$.
The effect of the turbulent diffusion has three contributions: 
\begin{equation}
F_{\mathcal{\eta}}=-2\eta_{T}\hat{\mathbf{r}}\cdot\left(\left\langle \boldsymbol{A}\right\rangle \times\left\langle \boldsymbol{J}\right\rangle \right)=-2\eta_{T}\left(\hat{\mathbf{r}}\cdot\nabla\right)\left(\left\langle \boldsymbol{A}\right\rangle \cdot\left\langle \boldsymbol{B}\right\rangle \right)+2\eta_{T}\left(\left\langle \boldsymbol{A}\right\rangle \cdot\nabla\right)\left\langle {B}\right\rangle_r +2\eta_{T}\left(\left\langle \boldsymbol{B}\right\rangle \cdot\boldsymbol{\nabla}\right)\left\langle {A}\right\rangle_r,\label{fluxEta}
\end{equation}
where the first term shows the same type of helicity flux as the diffusive
flux of turbulent magnetic helicity in Eq.\,(\ref{eq:fabp}).

\section{Results}

\subsection{Evolution of magnetic helicity fluxes and tilt and twist of emerging
BMR }

\begin{figure}
\centering \includegraphics[width=0.99\textwidth]{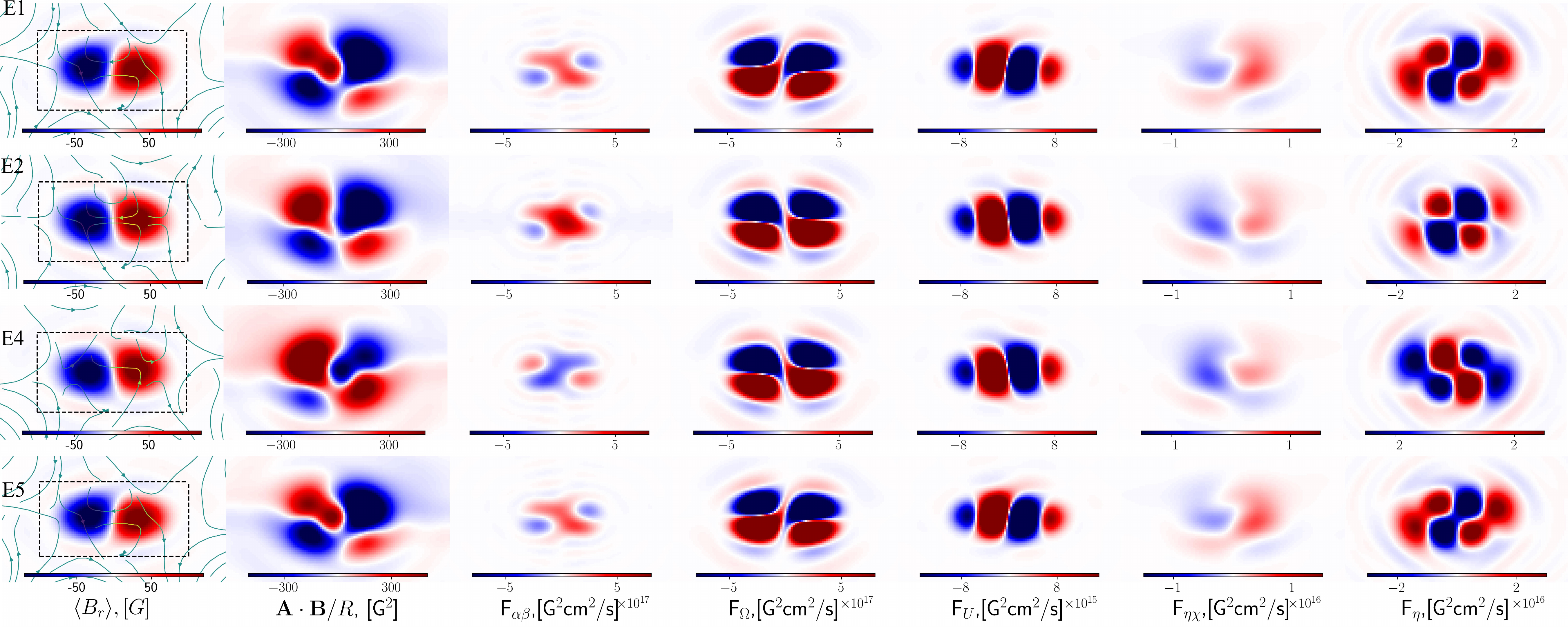}\caption{Snapshots of the magnetic field, the magnetic helicity density, and
the helicity density fluxes at the middle state of the active region
evolution: (a) the surface magnetic field; (b) the total magnetic
helicity density, $\boldsymbol{A\cdot B}$; the panels (c), (d), (e)
(f) and (g) show the density of the magnetic helicity flux distributions
$F_{\alpha\beta}$, $F_{\Omega}$, $F_{U}$, $F_{\eta\chi}$, and
$F_{\eta}$. The rectangle indicates the area used for calculating
the BMR helicity flux.}\label{fig3}
 
\end{figure}

Figure \ref{fig3} shows the snapshots of the magnetic field configuration
and helicity flux distributions after the BMR emergence for the cases
E1, E2, E4, and E5 at the middle stage of the BMR evolution. From
these model runs, we see that the magnitude and sign of the helicity
flux distribution significantly depends on the boundary conditions,
the sign of the tilt, and the mutual phase of the initial perturbations,
$\xi_{1}$ and $\xi_{2}$ in Eq.\,(\ref{eq:twtlt}). The runs with
the harmonic magnetic field boundary conditions, e.g., E2, E4, and
E5, show higher magnitudes of the surface magnetic helicity density
and the helicity flux initiated by the $\alpha$-effect, and magnetic
buoyancy, $F_{\alpha\beta}$, than the run E1. Similarly to the analysis
of \citet{PariDemBer2005AA}, we see that the emergence of the BMR
induces specific polarity patterns for each mechanism of the magnetic
helicity flux. The results for the helicity density flux distributions
due to the effects of the differential rotation and meridional circulations
are qualitatively in agreement with the patterns discussed in the
paper mentioned above. Also, we see that the flux, $F_{\alpha\beta}$,
which stems from the $\alpha$-effect and the helicity density initiated
by the BMR rise, is similar to the flux from the rotational motions
of the magnetic polarities relative to each other (see, \citealp{PariDemBer2005AA,Yamamoto2011PASJ}).

\begin{figure}
\centering \includegraphics[width=0.8\textwidth]{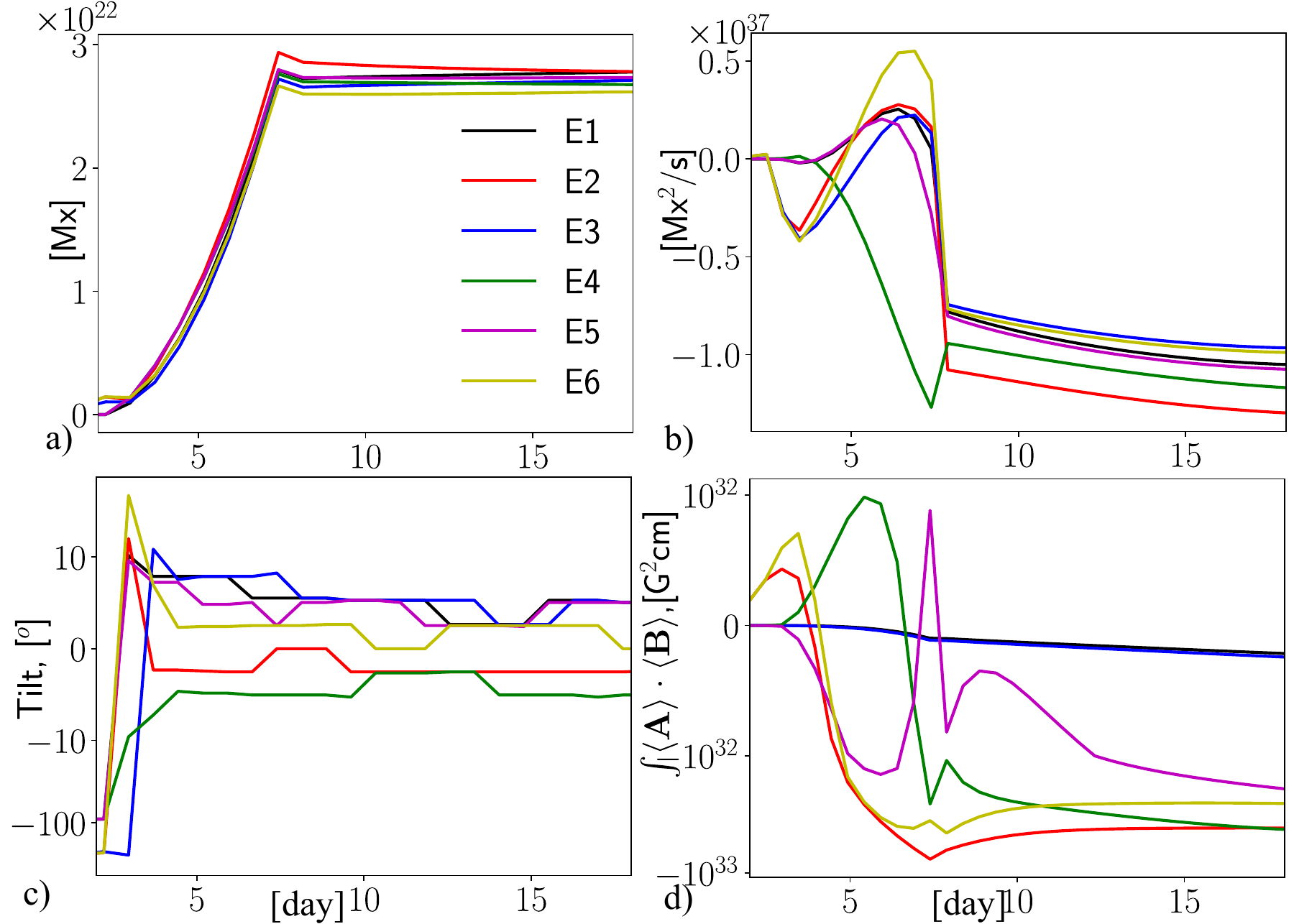} \caption{Evolution of the BMR's parameters during emergence: a) the total
unsigned radial magnetic field flux on from BMR's area marked by the
dash line in Figure~\ref{fig3}; b) the total helicity flux from
BMR; c) evolution of the BMR's tilt; d) the total surface magnetic
helicity density.}\label{fig4a}
 
\end{figure}

Figure \ref{fig4a} shows the evolution of the integral parameters:
the total unsigned radial magnetic field flux, the surface helicity
density, the total helicity flux, and the tilt of the BMR, calculated
following the procedure accepted in the SHARP routine \citep{Liuetal2014ApJ,Sun2024AA}.
The total magnetic flux and other integral parameters are calculated
in the area marked by the dashed line in Figure \ref{fig3}. Some
part of this flux further contributes to the large-scale dynamo in
the solar convection zone. It is noteworthy that restricting the area
of the integral neglects the effect of the large-scale, nonaxisymmetric
magnetic field in the helicity flux. The total helicity flux in cases
E2, E4, E5, and E6 sharply increases at the beginning of the BMR emergence
because of the helicity transport by the twisted magnetic field rising
from the convection zone. At the end of emergence, the largest contribution
to the helicity flux is due to the effect of differential rotation.
We find that if we integrate the flux over the whole surface, the
sharp increase of the flux, which is seen in Figure\,\ref{fig4a}b
at the beginning phase of BMR evolution, disappears. This tells about
the importance of the large-scale nonaxisymmetric magnetic field contribution
to the helicity flux. The BMR's tilt can evolve in a different way,
which depends on whether the BMR was twisted before emergence. For
example, for case E1, the BMR's tilt decreases continuously until
$\approx5{}^{\circ}$ and then it shows some variations about this
value caused by the BMR's evolution. The twisted BMRs show the anti-Hale
direction of the tilt at the beginning of their emergence.

The surface magnetic helicity density shows sign variations. For all
cases, except E4, we find the predominantly negative helicity density
of the magnetic field during the model runs. In the case of the negative
tilt, case E4, the BMR shows an inversion of the helicity sign from
positive to negative at the end of this model run. This is caused
by the differential rotation effect. Further results that support
this conclusion are shown in Figure \ref{fig5}. The runs, which employ
the harmonic boundary conditions, show a positive helicity density
at the very beginning of the BMR's emergence.

\begin{figure}
\centering \includegraphics[width=0.7\textwidth]{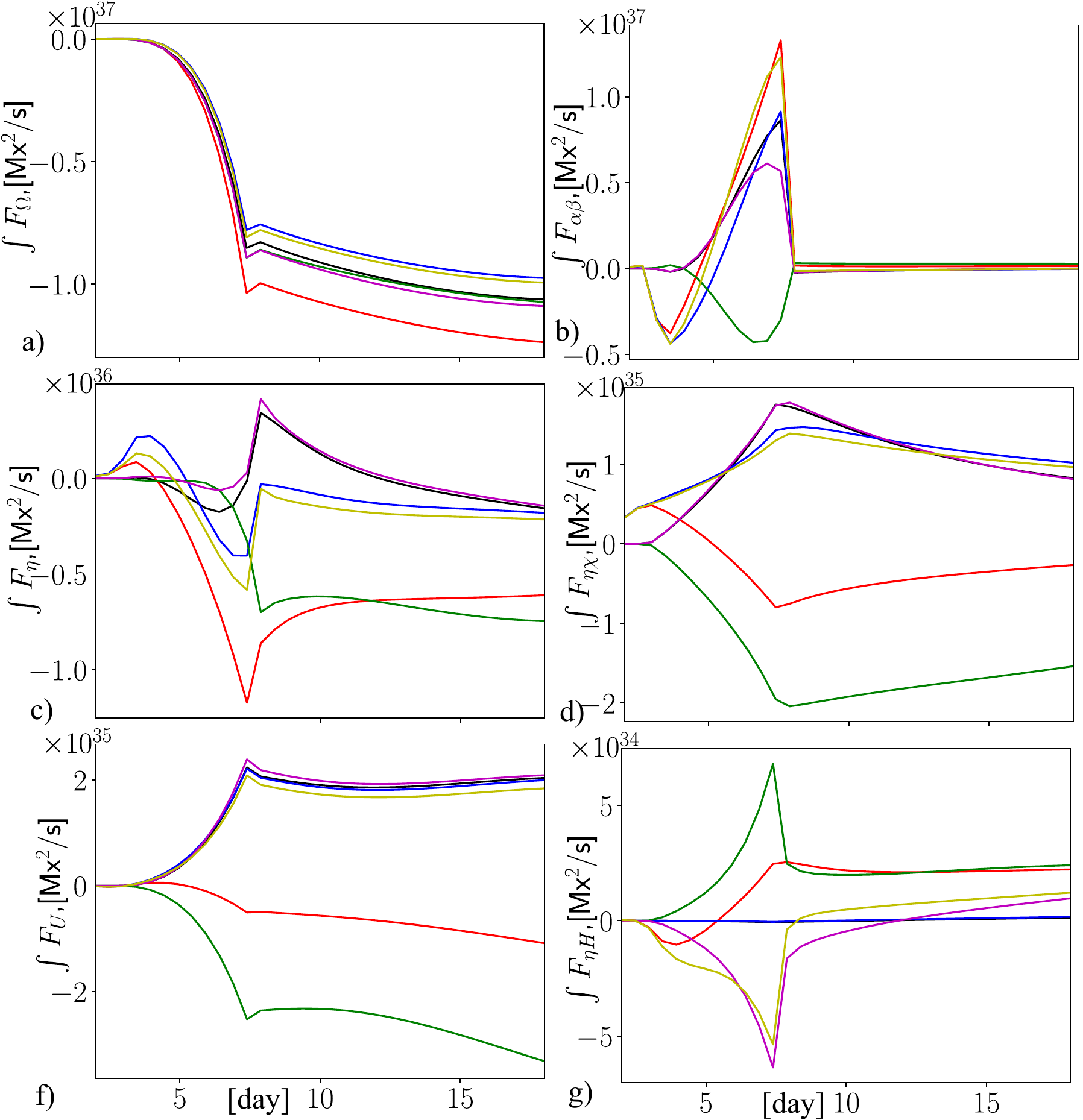} \caption{Evolution of the BMR's helicity flux during emergence: a) the evolution
of the helicity flux due to the differential rotation, $F_{\Omega}$;
b) the helicity flux by the BMRs' tilt/twist, $F_{\alpha\beta}$;
c) the flux initiated by BMRs' decay due to the turbulent diffusion,
$F_{\eta}$; d) the diffusive flux of the small-scale magnetic helicity;
e) the helicity flux by the meridional circulation, $F_{U}$; f) the
flux initiated by BMRs' decay due to the horizontal turbulent diffusion,
$F_{\eta H}$.}\label{fig4b}
 
\end{figure}

\begin{figure}
\centering \includegraphics[width=0.95\textwidth]{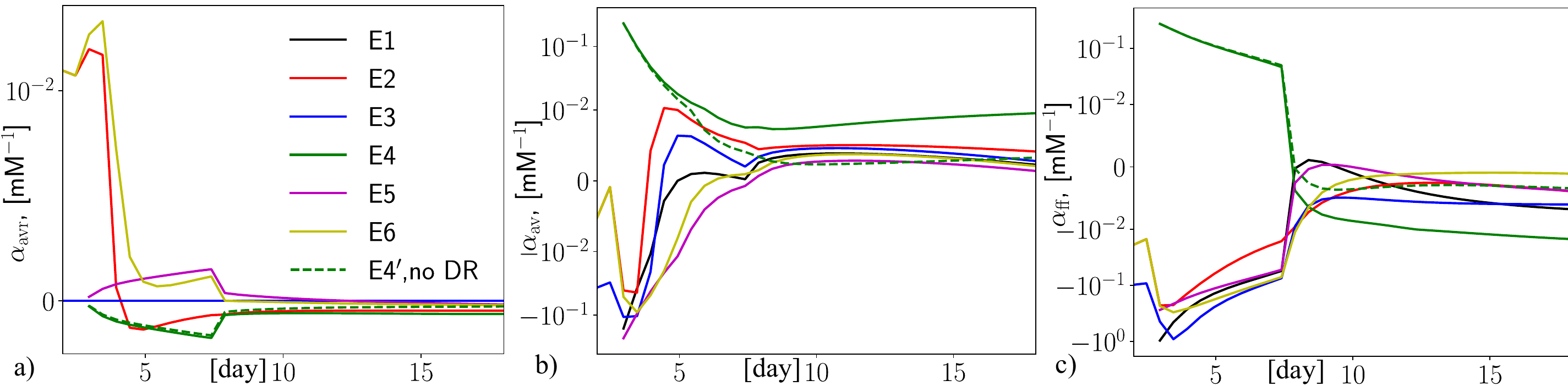} \caption{ Evolution of the parameters $\mathrm{\alpha}_{\mathrm{avr}}$ (a),
$\mathrm{\alpha}_{\mathrm{av}}$ (b), and $\mathrm{\alpha}_{\mathrm{ff}}$
(c). The green dashed line shows results for the run E4$^{\prime}$
with the neglected effect of the differential rotation.}\label{fig5}
\end{figure}

Figure \ref{fig4b} shows the evolution of the integral parameters
of the magnetic helicity flux. The helicity flux rate is higher than
the magnetic flux rate. This agrees with the analysis of observations
made by \citet{Liuetal2014ApJ,Nortoneal2023,Sun2024AA}. The interesting
finding is that the helicity flux due to the turbulent diffusion in
the radial direction, $F_{\eta V}$, dominates the contribution from
the diffusion in the horizontal direction $F_{\eta H}$. In other
words, in the expression of the turbulent diffusion helicity flux:
\begin{eqnarray}
F_{\mathcal{\eta}} & = & F_{\eta V}+F_{\eta H},\label{eq:Fh}\\
F_{\eta V} & = & -2\eta_{T}\nabla\left(\left\langle \boldsymbol{A}\right\rangle \cdot\left\langle \boldsymbol{B}\right\rangle \right)+2\eta_{T}\left(\left\langle \boldsymbol{B}\right\rangle \cdot\nabla\right)\left\langle \boldsymbol{A}\right\rangle ,\label{Fev}\\
F_{\eta H} & = & 2\eta_{T}\left(\left\langle \boldsymbol{A}\right\rangle \cdot\boldsymbol{\nabla}\right)\left\langle \boldsymbol{B}\right\rangle ,\label{FeH}
\end{eqnarray}
we should take the radial components of the fluxes into account. Using
the identity $\hat{\mathbf{r}}\cdot\left\langle \boldsymbol{A}\right\rangle ^{(p)}=0$,
we see that the second term in $F_{\eta V}$ is zero. The contribution
$F_{\eta V}$ dominates $F_{\eta H}$. Moreover, because of the condition
$\nabla\cdot\langle\mathbf{A}\rangle^{(p)}=0$, the total surface
integral of contribution $F_{\eta H}$ is close to zero. The comparison
of Figures \ref{fig4b}(c) and(g) shows that $F_{\eta V}$ is approximately
4 orders of magnitude higher than $F_{\eta H}$.

Comparing Figures \ref{fig4b} (c) and (d), we see that the diffusive
decay of the BMR is an order of magnitude larger than the diffusive
flux of the small-scale magnetic helicity (Eq.~\ref{eq:hdf}). This
is because the magnitude of the turbulent diffusion of the magnetic
field is by an order of magnitude larger than the turbulent diffusion.

We compute the distributions of the current helicity density $\left\langle \mathbf{B}_{\|}\right\rangle \left(\nabla\times\left\langle \boldsymbol{B}\right\rangle \right)_{\|}$
and the total current helicity. Here $\langle\mathbf{B}_{\|}\rangle=\sin\theta\langle B_{r}\rangle+\cos\theta\langle B_{\theta}\rangle$
is the line-of-sight magnetic field. For the potential boundary conditions,
the radial component of current is zero, $\left(\nabla\times\left\langle \boldsymbol{B}\right\rangle \right)_{r}=0$.
Nevertheless, the projection effects can result in nonzero values
of $\left\langle \boldsymbol{B}_{\|}\right\rangle \left(\nabla\times\left\langle \boldsymbol{B}\right\rangle \right)_{\|}$.
Following \citet{Hagino2004}, we calculated the weighted values of
the force-free parameter $\alpha$, using the average over the active
region amplitudes of the electric currents and magnetic field: 
\begin{eqnarray}
\mathrm{\alpha}_{\mathrm{av}} & = & \frac{\int\left\langle B_{\|}\right\rangle \left(\nabla\times\left\langle \boldsymbol{B}\right\rangle \right)_{\|}\mathrm{d}S}{\int\left\langle B_{\|}\right\rangle ^{2}\mathrm{d}S},\label{affp}\\
\mathrm{\alpha}_{\mathrm{ff}} & = & \frac{\int\left\langle \boldsymbol{B}\right\rangle \cdot\left(\nabla\times\left\langle \boldsymbol{B}\right\rangle \right)\mathrm{d}S}{\int\left\langle B\right\rangle ^{2}\mathrm{d}S},\label{eq:aff}\\
\mathrm{\alpha}_{\mathrm{avr}} & = & \frac{\int\left\langle B_{r}\right\rangle \left(\nabla\times\left\langle \boldsymbol{B}\right\rangle \right)_{r}\mathrm{d}S}{\int\left\langle B_{r}\right\rangle ^{2}\mathrm{d}S},\label{affr}
\end{eqnarray}
Figure \ref{fig5} shows the results. The maximum amplitude of the
twist parameters, $\mathrm{\alpha}_{avr}$, $\mathrm{\alpha}_{av}$
and the total twist parameter $\mathrm{\alpha}_{\mathrm{ff}}$, are
by an order of magnitude larger than $\alpha_{\mathrm{best}}$ in
observations \citep[e.g.,][]{Tian2005SoPh,Kuzanyan2006}. The state-of-the
convective magnetic flux emergence simulations of \citet{Toriumi2024}
for a kink-unstable magnetic tube showed the same magnitude of $\mathrm{\alpha}_{\mathrm{ff}}$
as in our models. 
We made the additional run for the setup E4, where we skipped the
effect of the differential rotation on the magnetic field evolution
(case E4$^{\prime}$ in Figure \ref{fig5}). It shows that shortly
after emergence (after day 5), the magnitude and sign of $\mathrm{\alpha}_{av}$
are determined by the effect of the differential rotation on the meridional
component of the magnetic field. 
The runs with the potential boundary conditions show $\mathrm{\alpha}_{\mathrm{ff}}\gg\mathrm{\alpha}_{\mathrm{av}}$
because $\left(\nabla\times\left\langle \boldsymbol{B}\right\rangle \right)_{r}=0$.
This means that in such a situation, the twist of the magnetic field
in the horizontal direction dominates the twist in the vertical direction.

\subsection{The hemispheric helicity rule}

\begin{figure}
\centering \includegraphics[width=0.75\textwidth]{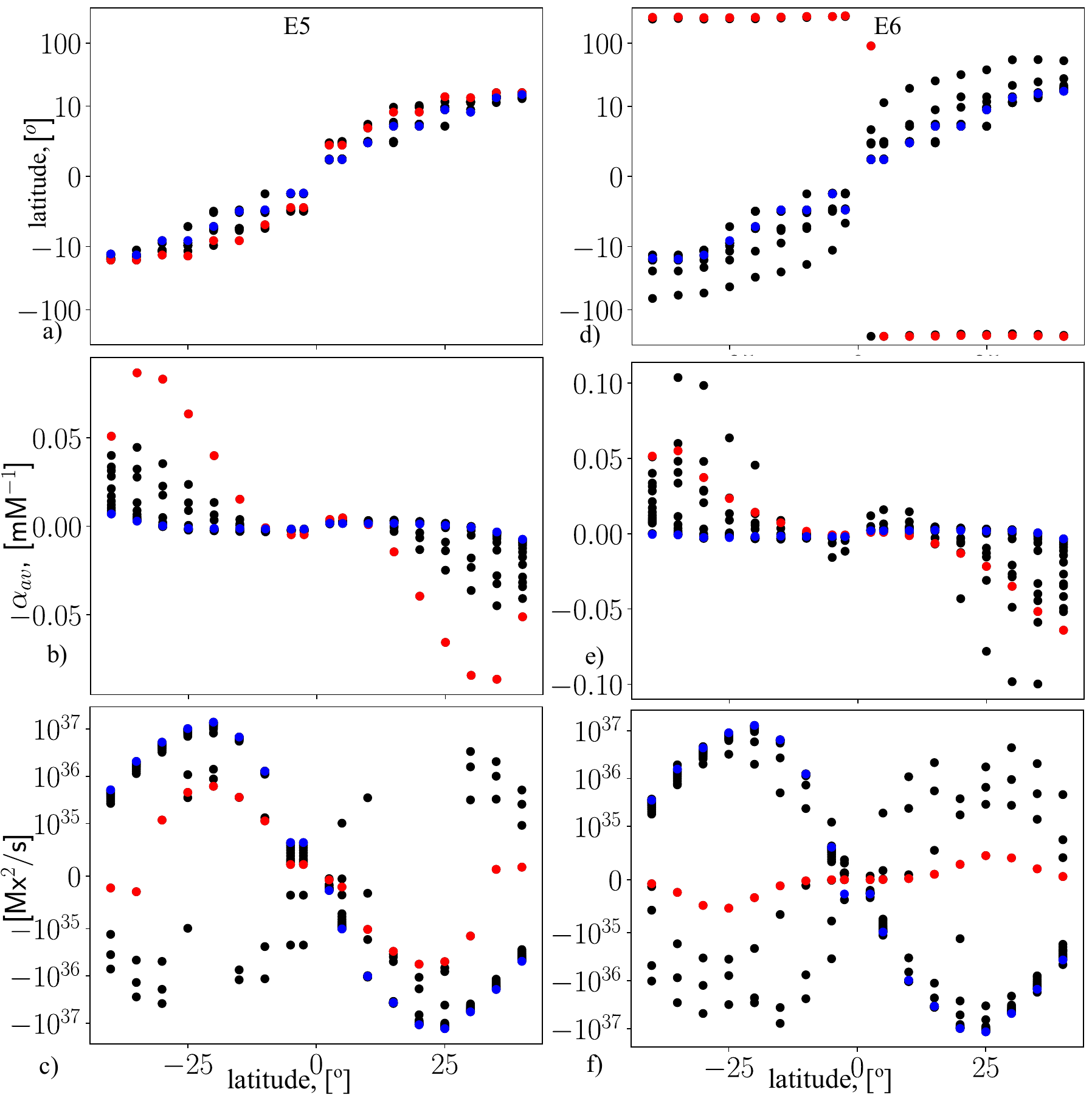} \caption{Latitudinal dependence of the tilt (a), the parameter $\mathrm{\alpha}_{\mathrm{ff}\|}$(b)
and the total flux (c) for the model setup E5. The red circles show
the value at the beginning of the BMR emergence, and the blue circles
show the same for the final stage of the run. The second column with
panels d), e), and f) shows the same parameters for the model setup
E6. The vertical scale in panel (d) is linear in the range of $\pm10^{\circ}$
and logarithmic outside this range.}\label{fig6}
 
\end{figure}

We compute the latitudinal variations of the current helicity parameter
$\mathrm{\alpha}_{av}$ and the total helicity flux to see how well
the theoretically studied active regions fit into the hemispheric
helicity rule. Figure \ref{fig6} shows these parameters together
with the latitudinal variation of tilt for the E5 and E6 model setups. 
Model E6 shows an angle of inclination about twice that of the tilt
angle profile of model E5. In addition, the scatter of the latitudinal
profiles of $\alpha_{av}$ and the total helicity flux is higher in model E6. We find that for these types of active regions, the hemispheric rule changes during evolution. In the final state, the hemispheric rule is determined by the effect of the differential rotation.

\begin{figure}
\centering \includegraphics[width=0.75\textwidth]{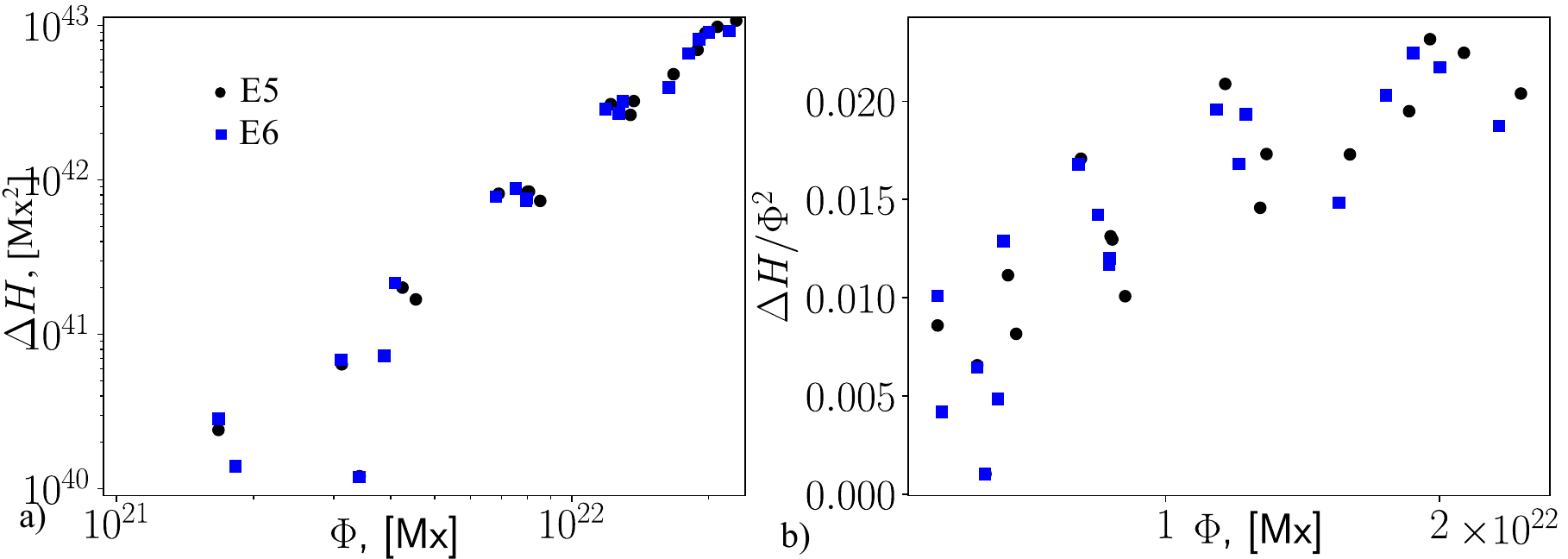} \caption{a) The helicity $\Delta H=\protect\int\partial_{t}H\mathrm{d}t$
accumulated in the BMRs versus the maximum total flux of the radial
magnetic field, $\Phi$; the black circles show results for the model
setup E5, and blue squares show the same for the model setup E6; b)
the same for the normalized value, $\Delta H/\Phi$.}\label{fig7}
 
\end{figure}

Using the helicity flux calculations, we compute the total helicity
that is transported through the area of the BMRs marked by a rectangle
in Figure\,\ref{fig3} \citep{Toriumi2024,Sun2024AA}, 
\begin{equation}
\Delta H=\int\frac{\partial H}{\partial t}\mathrm{d}t
\end{equation}
The data drawn in Figure \ref{fig7}a show a power-law dependence,
$\Delta H\sim0.02\Phi^{2}$, where $\Phi$ the maximum total flux
of the radial magnetic field. The results of \citet{Sun2024AA} suggested
a similar power exponent. Such a value of the power exponent shows
that the magnetic field configuration is topologically close to a simple linkage of two close magnetic loops around each other. {In contrast to \citet{Sun2024AA}, our
model shows some increase in the linkage parameter with the
amount of magnetic flux. This difference is probably because of the different definitions for the helicity flux. The paper of \citet{Sun2024AA} employs the relative magnetic helicity.}

\section{Discussion and conclusions}
In this study, we modeled some typical configurations of magnetic
bipolar active regions (BMR) using the previously developed 3D non-linear
mean-field MHD solar dynamo model, which includes the emergence of BMRs
due to a magnetic buoyancy instability \citep{PKT23}. In this model,
the turbulent hydrodynamic and magnetic helicity (the $\alpha$-effect),
acting locally on the unstable parts of the toroidal magnetic field
that form the bipolar active region, produces the twist and tilt of
the magnetic field inside the BMRs. We calculated the magnetic helicity
flux from the dynamo region to the outer layers initiated by the BMR's
emergence. Starting from \citet{Fan2001ApJ}, similar studies were
done previously using simulations of the kink-unstable magnetic flux
tube in the convective media in a number of papers (see, e.g.,\citealp{Prior2014ApJ,Prior2019JP,Toriumi2019,Toriumi2024}).
Similarly to these papers, in our simulations, we did not take into
account the non-axisymmetric hydrodynamic motions and heat transport
around the BMR. However, the dynamo model describes the non-linear
magnetic effects on the axisymmetric flow and the convective heat
transport. Our approach to modeling the evolution of the photospheric
BMR is rather simple. Nevertheless, this model is a step toward a
consistent picture of the large-scale convection zone dynamo describing
important parameters of the active regions on the solar surface such
as the tilt and twist of the magnetic field. 

In the standard surface flux-transport model (hereafter SFT, e.g.,
\citealp{Yeates2023a}), a solar active region is approximated by
a simple bipolar magnetic structure, which is tilted according to
Joy's law. In SFT models, the hemispheric helicity rule can result
from the differential rotation effect on the surface evolution of
BMRs \citep{Prior2014ApJ}. \citet{Hawkes2019AA} utilized the SFT
to estimate the helicity flux initiated by the emergence and evolution
of the BMRs. Their expression of the helicity flux employs the effects
of the global flow, including the differential rotation and meridional
circulation, and the effect of the surface turbulent diffusion. The
mean-field formalism, which is employed here to model both the large-scale
dynamo and the BMRs, allows us to estimate the possible effect of
the tilt and twist of the rising active regions on the helicity flux
and evolution of the mean twist of the solar active regions. We also
take into account the radial structure of the magnetic field below
the surface. For the first time, such calculations are performed using
a large-scale 3D solar dynamo model, consistent with helioseismic
and surface observations \citep{PKT23,Manda2024}.

We see that the action of the $\alpha$-effect on the rising part
of the large-scale toroidal magnetic field can result in a complicated
evolution of the magnetic flux in the emerging BMRs. Similarly to
the numerical model of the twisted magnetic tubes subjected to the
kink instability \citep{Fan2001ApJ}, our model, e.g., the runs E2,
E3, and E6, show a rotation of the BMR when the $\alpha$-effect
is applied before the rising stage of the BMR evolution. Rotation
of BMRs is often observed in the evolution of the solar active regions
\citep{Tian2005SoPh}.

The model shows a monotonic increase in the magnetic flux before the
end of the emerging stage. The helicity flux can grow sharply at the
beginning because of the rise of the twisted magnetic field. This
agrees with the results of the observations of \citet{Liuetal2014ApJ,Sun2024AA}
and with the state-of-the-art model of \citet{Toriumi2024}. Similarly
to the observational results, we find that at the end of the BMR evolution,
the major effect in the helicity flux is due to the differential rotation.
Another interesting finding is that our model shows a significant
helicity flux induced by the radial gradient of the magnetic helicity
of the BMRs. This flux was theoretically suggested by \citet{Mitra2011,Kleeorin2022,Subramanian2023ApJ}.
The mean-field models showed that it can significantly affect the
dynamo solution \citep{Guerrero2010}. Here, we calculated this contribution
to the helicity flux for the BMR. We plan to study its effect on the
dynamo cycles in our future studies.

The model shows that, unlike the magnetic helicity flux, the twist
parameters, such as $\alpha_{avr}$ and $\alpha_{av}$ (Eqs~\ref{affp}\,--\,\ref{affr})
can develop quickly during the initial phase of BMR emergence. This
reflects both the effect of the current preexisting before the magnetic
field emerges at the surface and the evolution of the magnetic field
topology inside the BMR during the rising phase. In our model, the
result of this evolution depends on the initial and surface boundary
conditions. In observations \citep[e.g.,][]{Leka1996ApJ}, this behavior
is often interpreted as a sign that the magnetic field is twisted
before it emerges. In runs E2 and E6 (Table 1), the initial positive
sign of $\alpha_{avr}$ and $\alpha_{av}$ corresponds to the sign
of the mean electromotive force applied to the toroidal magnetic field
before the BMR rise. These parameters quickly change to negative (Figure
\ref{fig4a}b) because of the dynamic response of the magnetic configuration
to the conservation of total magnetic helicity. This result is consistent
with interpretations of the helicity of the solar active regions suggested
by the mean-field dynamo models \citep{Sokoloff2006}. Yet, our results
show that after the BMR emergence, the helicity of the solar active
regions is quickly modified by the effect of the differential rotation.
In the final stage, the BMR's twist parameter $\alpha_{av}$ shows
the hemispheric helicity rule. In the model, the small magnitude of
twist is supported by the effect of differential rotation. It is noteworthy
that the hemispheric helicity rule can depend on the phase of the
magnetic cycle \citep[e.g.,][]{Zhang2010}. Here, we do not consider
this effect. The helicity flux shows the hemispheric helicity rule
as well. Our results agree with the results of \citet{Berger2000}
and \citet{Hawkes2019AA} in this regard.

{In this paper, we discussed the helicity flux from the dynamo region to the corona caused by the emergence of individual BMR. For symmetric BMR emergence, the surface integrals of the radial helicity fluxes, as well as the volume integrals of the helicity density over the solar convection zone and corona, is        identically zero \citep{Berger2000}. Figure \ref{fig6} also suggests the same result. The injection of the BMRs that is asymmetric about the equator can produce the helicity flux across the equator. The example of this can be seen in the magnetic helicity distribution presented in Figure \ref{fig1}c. The turbulent electromotive force, that is, the term $\left(\boldsymbol{\mathcal{E}}\times\left\langle \boldsymbol{A}\right\rangle \right)$, can produce the cross-equatorial helicity flux as well.  Such fluxes can affect the parity of the dynamo solution \citep{Mitra2010}. We postpone discussion of this effect to another study. }

Let us summarize our findings. Using the mean-field MHD formalism,
we estimated the influence of BMR's tilt and twist on the helicity
flux. Our model shows that the action of the $\alpha$-effect before
and during the BMR emergence results in a complex evolution of magnetic
configuration, affecting variations of the twist and tilt of the emergent
BMR. Such variations are caused by the magnetic helicity conservation,
large-scale flows, and turbulent diffusion of the magnetic field.
The findings highlight the differential rotation as a key driver of
helicity flux, with significant effects induced by radial magnetic
helicity gradients. While the twist parameters evolve quickly during
the BMR emergence, influenced by initial and surface conditions, the
differential rotation strongly impacts helicity flux consistency with
the hemispheric helicity rule. We conclude that BMR twists and helicity
quickly adapt to the post-emergence state. Further research is needed
to understand the impacts of active region helicity fluxes on the
dynamo cycles.
\begin{flushleft}
\textit{Acknowledgments}: Resources supporting this work were provided
by the NASA High-End Computing (HEC) Program through the NASA Advanced
Supercomputing (NAS) Division at Ames Research Center. This work is
supported by the Chinese Academy of Sciences (CAS) President's International
Fellowship Initiative (PIFI) No 2024PVA0093. In addition, VP thanks the financial
support of the Ministry of Science and Higher Education (subsidy No.075-GZ/C3569/278). 
\par\end{flushleft}

\appendix{}

\section{Magnetic helicity budget in the dynamo model.}\label{A}

The details of our model can be found in the paper of \citet{PKT23}.
Here, we review the parts of the model that are directly related to
the subject of the paper.

The dynamo model employs the mean electromotive force, $\boldsymbol{\mathcal{E}}$, as follows. 
\begin{equation}
\mathcal{E}_{i}=\left(\alpha_{ij}+\gamma_{ij}\right)\left\langle B\right\rangle _{j}-\eta_{ijk}\nabla_{j}\left\langle B\right\rangle _{k}+\mathcal{E}_{i}^{\mathrm{(BMR)}},\label{eq:emf}
\end{equation}
where $\boldsymbol{\mathcal{E}}^{\mathrm{(BMR)}}$ represents the
contribution, which prescribes the generation of the bipolar active
regions, $\alpha_{ij}$ describes the turbulent generation by the
hydrodynamic and magnetic helicity, $\gamma_{ij}$ is the turbulent
pumping and $\eta_{ijk}$ - the eddy magnetic diffusivity tensor.
In particular, the $\alpha$-effect tensor includes the effect of magnetic helicity conservation (\citealp{Kleeorin1982,Kleeorin1999}),
\begin{eqnarray}
\alpha_{ij} & = & C_{\alpha}\psi_{\alpha}(\beta)\alpha_{ij}^{{\rm K}}+\alpha_{ij}^{{\rm M}}\psi_{\alpha}(\beta)\frac{\left\langle \mathbf{a}\cdot\mathbf{b}\right\rangle \tau_{c}}{4\pi\overline{\rho}\ell_{c}^{2}}.\label{alp2d}
\end{eqnarray}

Here $C_{\alpha}$ is the dynamo parameter characterizing the magnitude
of the hydrodynamic $\alpha$-effect, and $\alpha_{ij}^{{\rm K}}$
and $\alpha_{ij}^{{\rm M}}$ describe the anisotropic properties of
the kinetic and magnetic $\alpha$-effect (\citealp{Pipin2008a,PK19,BRetal23}).
The radial profiles of $\alpha_{ij}^{H}$ and $\alpha_{ij}^{M}$ depend
on the mean density stratification and the spatial profiles of the
convective velocity $u_{c}$, and on the Coriolis number, 
\begin{equation}
{\rm Co}=2\Omega_{0}\tau_{c},\label{eq_M8}
\end{equation}
where $\Omega_{0}$ is the global angular velocity of the star, and
$\tau_{c}$ is the convective turnover time. The magnetic quenching
function $\psi_{\alpha}(\beta)$ depends on the parameter $\beta=|\langle\mathbf{B}\rangle|/\sqrt{4\pi\overline{\rho}u_{c}^{2}}$
\citep{PK19}. We used the analytical expressions of the coefficients
of $\mathbf{\boldsymbol{\mathbf{\mathcal{E}}}}$ given by \citet{Pipin2008a}.
The initiation of the bipolar magnetic regions is determined by $\mathcal{E}_{i}^{(\mathrm{BMR})}$,
see Section \ref{subsec:Model} and for more details in \citet{PKT23}:
\begin{equation}
\boldsymbol{\mathcal{E}}^{(\mathrm{BMR})}=\alpha_{\beta}^{\rm BMR}\left\langle \boldsymbol{B}\right\rangle +V_{\beta}\left(\hat{\boldsymbol{r}}\times\left\langle \mathbf{B}\right\rangle \right),\label{eq:ebmr}
\end{equation}
where the first term takes into account the BMR's tilt/twist and the
second term models the emergence of the surface magnetic region in
the bipolar form. The induction equation (Eq.~\ref{eq:MFE}) describes
the evolution of both the large-scale magnetic field and
the evolution of the BMRs. {It is noteworthy that the longitudinal averaging of $\boldsymbol{\mathcal{E}}^{(\mathrm{BMR})}$ results in the additional generation effect of the large-scale axisymmetric magnetic field and the additional sources of the magnetic flux loss \cite{P22}. Therefore, the critical threshold of the mean field parameter $C_{\alpha}$ decreases in the presence of $\boldsymbol{\mathcal{E}}^{(\mathrm{BMR})}$. Presumably, the mean-field dynamo can operate with the BMR electromotive force $\boldsymbol{\mathcal{E}}^{(\mathrm{BMR})}$ alone, starting from a quite large amount of toroidal magnetic flux inside the convection zone, i.e., due to the non-linear dynamo instability \citep{Ferriz1994}. Such dynamo instability can depend on the parameters of the BMR's injection functions, see Appendix \ref{C}.  However, we have not studied this issue.}

Uncurling the induction equation for $\left\langle \boldsymbol{B}\right\rangle $,
we obtain the evolution equation for the mean vector-potential, 
\begin{equation}
\frac{\partial\left\langle \boldsymbol{A}\right\rangle }{\partial t}=\left(\boldsymbol{\mathcal{E}}+\left\langle \boldsymbol{U}\right\rangle \times\left\langle \boldsymbol{B}\right\rangle \right)+\boldsymbol{\nabla}h,\label{eq:mfA}
\end{equation}
where $h$ is an arbitrary scalar function.

Before proceeding further, we discuss the problem of the gauge. We
decompose the large-scale magnetic field and flows into the sum of
axisymmetric and nonaxisymmetric parts: 
\begin{eqnarray*}
\left\langle \boldsymbol{B}\right\rangle  & = & \left\langle \boldsymbol{\overline{B}}\right\rangle +\left\langle \tilde{\boldsymbol{B}}\right\rangle ,\\
\left\langle \boldsymbol{A}\right\rangle  & = & \left\langle \boldsymbol{\overline{A}}\right\rangle +\left\langle \tilde{\boldsymbol{A}}\right\rangle ,
\end{eqnarray*}
Moreover, the nonaxisymmetric part is decomposed into a sum of the
poloidal and toroidal superpotentials: 
\begin{equation}
\left\langle \tilde{\boldsymbol{B}}\right\rangle =\nabla\times\boldsymbol{r}\tilde{T}\left(\boldsymbol{r},t\right)+\nabla\times\nabla\times\boldsymbol{r}\tilde{S}\left(\boldsymbol{r},t\right),\label{eq:nax}
\end{equation}
where $\boldsymbol{r}$ is the radius vector in the spherical coordinate
system; $\tilde{S}$ and $\tilde{T}$ are the scalar potentials\citep{Krause1980}.
It should be noted that the axisymmetric field can be decomposed into
the sum of the poloidal and toroidal parts as well: 
\begin{equation}
\left\langle \boldsymbol{\overline{B}}\right\rangle =B\boldsymbol{e}_{\phi}+\nabla\times\left(\frac{A\boldsymbol{e}_{\phi}}{r\sin\theta}\right)\label{eq:ax}
\end{equation}
where the scalars $B$ and $A$ are the functions of time, $r$ is
the radius, $\theta$ is the co-latitude (the polar angle), and $\boldsymbol{e}_{\phi}$
is the unit vector along the azimuth. In our notations, we can write,
\begin{eqnarray*}
\left\langle \boldsymbol{\overline{A}}\right\rangle  & \equiv & \frac{A}{r\sin\theta}\boldsymbol{e}_{\phi}\\
\left\langle \tilde{\boldsymbol{A}}\right\rangle  & = & \boldsymbol{r}\tilde{T}\left(\boldsymbol{r},t\right)+\nabla\times\boldsymbol{r}\tilde{S}\left(\boldsymbol{r},t\right)+\boldsymbol{\nabla}g,
\end{eqnarray*}

Following \citet{Krause1980}, we note that the arbitrarily chosen
scalar $h$ is a function of the radial coordinate, $r$, and the
same is true for $g$. This uncertainty can be removed if we consider
the integrals of the scalars $\tilde{T}$ and $\tilde{S}$ over the
solid angle normalized to zero, i.e. $\int_{0}^{2\pi}\int_{-1}^{1}\tilde{S}\mathrm{d}\mu\mathrm{d}\phi=\int_{0}^{2\pi}\int_{-1}^{1}\tilde{T}\mathrm{d}\mu\mathrm{d}\phi=0$,
where $\mu=\cos\theta$. This gauge is valid by default for $\left\langle \boldsymbol{\overline{A}}\right\rangle $
because it satisfies the Coulomb gauge, $\boldsymbol{\nabla}\cdot\left\langle \boldsymbol{\overline{A}}\right\rangle \equiv0$. {It is noteworthy that  the scalars $\tilde{T}$ and $\tilde{S}$ are super-potentials of the pure nonaxisymmetric magnetic field. To solve the induction equation for the nonaxisymmetric magnetic field, we employ the spherical harmonic decomposition (hereafter SHD) library of \cite{shtns}  with $\ell_{max}=48$.  The solid angle integrals of the SHD of $\tilde{T}$ and $\tilde{S}$  are identically zero. Also the horizontal divergence of the vector potential $\left\langle \tilde{\boldsymbol{A}}\right\rangle$ is zero as well, $\boldsymbol{\nabla}_{H}\cdot \left\langle \tilde{\boldsymbol{A}}\right\rangle=0$ \citep{Berger2018}. } 
Assuming the above normalization procedure is valid and decompositions of the
large-scale field given by Equations (\ref{eq:nax}) and (\ref{eq:ax}),
we can omit the contribution $\boldsymbol{\nabla}h$ from the equation
for the mean vector-potential evolution. After some algebra, we get
the evolution equation for the helicity density of the mean magnetic
field, 
\begin{eqnarray}
\mathrm{\frac{\partial\left\langle \boldsymbol{A}\right\rangle \cdot\left\langle \boldsymbol{B}\right\rangle }{\partial t}}+\boldsymbol{\nabla}\cdot\left(\left\langle \boldsymbol{U}\right\rangle \left\langle \boldsymbol{A}\right\rangle \cdot\left\langle \boldsymbol{B}\right\rangle \right) & = & 2\boldsymbol{\mathcal{E}}\cdot\left\langle \boldsymbol{B}\right\rangle +2\nabla\cdot\left(\boldsymbol{\mathcal{E}}\times\left\langle \boldsymbol{A}\right\rangle \right)\label{eq:AB}\\
 & + & 2\nabla\cdot\left\langle \boldsymbol{B}\right\rangle \left(\left\langle \boldsymbol{A}\right\rangle \cdot\left\langle \boldsymbol{U}\right\rangle \right)-\nabla\cdot\left\langle \boldsymbol{U}\right\rangle \left(\left\langle \boldsymbol{A}\right\rangle \cdot\left\langle \boldsymbol{B}\right\rangle \right)\nonumber \\
 & - & 2\eta\left\langle \boldsymbol{B}\right\rangle \cdot\left\langle \boldsymbol{J}\right\rangle +\nabla\cdot\left(2\eta\left\langle \boldsymbol{A}\right\rangle \times\left\langle \boldsymbol{J}\right\rangle +\left\langle \boldsymbol{A}\right\rangle \times\frac{\partial\left\langle \boldsymbol{A}\right\rangle }{\partial t}\right)
\end{eqnarray}
where we use the following identities 
\begin{equation}
\frac{\partial\left\langle \boldsymbol{A}\right\rangle \cdot\left\langle \boldsymbol{B}\right\rangle }{\partial t}=\left\langle \boldsymbol{A}\right\rangle \cdot\frac{\partial\left\langle \boldsymbol{B}\right\rangle }{\partial t}+\nabla\times\left\langle \boldsymbol{A}\right\rangle \cdot\frac{\partial\left\langle \boldsymbol{A}\right\rangle }{\partial t}=2\left\langle \boldsymbol{A}\right\rangle \cdot\frac{\partial\left\langle \boldsymbol{B}\right\rangle }{\partial t}+\nabla\cdot\left(\left\langle \boldsymbol{A}\right\rangle \times\frac{\partial\left\langle \boldsymbol{A}\right\rangle }{\partial t}\right),\label{eq:ABa}
\end{equation}

Subtracting Equation~(\ref{eq:AB}) from the total magnetic helicity
balance equation (Eq.~\ref{eq:intcons}), we get 
\begin{eqnarray}
\frac{\mathrm{d}}{\mathrm{d}t}\int\left\langle \boldsymbol{a\cdot}\boldsymbol{b}\right\rangle dV & = & -2\int\left(\boldsymbol{\mathcal{E}}\cdot\left\langle \boldsymbol{B}\right\rangle \right)dV-\int\frac{\left\langle \boldsymbol{a\cdot}\boldsymbol{b}\right\rangle }{R_{m}\tau_{c}}dV\label{eq:ab-2}\\
 & - & \oint d\boldsymbol{S}\cdot\boldsymbol{F}^{\left\langle ab\right\rangle }+\oint d\boldsymbol{S}\cdot\left\langle \boldsymbol{U}\right\rangle \left(\left\langle \boldsymbol{A}\right\rangle \cdot\left\langle \boldsymbol{B}\right\rangle \right)\nonumber \\
 & -2 & \oint d\boldsymbol{S}\cdot\left(\boldsymbol{\mathcal{E}}\times\left\langle \boldsymbol{A}\right\rangle \right)-2\oint d\boldsymbol{S}\cdot\left\langle \boldsymbol{B}\right\rangle \left(\left\langle \boldsymbol{A}\right\rangle \cdot\left\langle \boldsymbol{U}\right\rangle \right).\nonumber 
\end{eqnarray}
Its differential form reads
\begin{eqnarray}
\frac{\mathrm{d}}{\mathrm{d}t}\left\langle \boldsymbol{a\cdot}\boldsymbol{b}\right\rangle  & = & -2\boldsymbol{\mathcal{E}}\cdot\left\langle \boldsymbol{B}\right\rangle -\frac{\left\langle \boldsymbol{a\cdot}\boldsymbol{b}\right\rangle }{R_{m}\tau_{c}}-\nabla\cdot\boldsymbol{F}^{\left\langle ab\right\rangle }+\nabla\cdot\left\langle \boldsymbol{U}\right\rangle \left(\left\langle \boldsymbol{A}\right\rangle \cdot\left\langle \boldsymbol{B}\right\rangle \right)\label{eq:ab}\\
 & - & 2\nabla\cdot\left(\boldsymbol{\mathcal{E}}\times\left\langle \boldsymbol{A}\right\rangle \right)-2\nabla\cdot\left\langle \boldsymbol{B}\right\rangle \left(\left\langle \boldsymbol{A}\right\rangle \cdot\left\langle \boldsymbol{U}\right\rangle \right).\nonumber 
\end{eqnarray}

This equation shows that the large-scale dynamo produces the magnetic
helicity in the bulk of the convection zone by means of the turbulent
electromotive force, i.e., due to the source term $-\left(\boldsymbol{\mathcal{E}}\cdot\left\langle \boldsymbol{B}\right\rangle \right)$.
This term potentially leads to the so-called catastrophic quenching
problem because of the magnetic helicity contribution to the $\alpha$-effect
\citep{Frisch1975}. The other terms of Equation~(\ref{eq:ab}) describe
the decay of the magnetic helicity due to the microscopic diffusion
and the helicity fluxes because of the turbulent processes, i.e.,
$\boldsymbol{F}^{\left\langle ab\right\rangle }$, and due to effects
of the large-scale dynamo evolution. From previous studies, we know
that the turbulent fluxes of the small-scale magnetic helicity alleviate
the catastrophic quenching of the $\alpha$-effect.

To calculate the helicity fluxes at the top of the dynamo domain, we employ  the isotropic expression for the turbulent diffusion, i.e., $\eta_{ijk}=\eta_{T}\varepsilon_{ijk}\nabla_{j}$,
where $\varepsilon_{ijk}$ is fully antisymmetric Levi-Chevita symbol, and for the $alpha$ effect as well.
For presentation, we denote the different contributions of the helicity
flux density as follows:
\begin{eqnarray}
F_{H} & = &F_{\Omega}+F_{U}+F^{\left\langle ab\right\rangle }  +F_{\mathcal{E}},\label{eq:EH}
\end{eqnarray}
and further decompose $F_{\mathcal{E}}=F_{\mathcal{\alpha}\beta}+F_{\mathcal{\eta}}$, where 
\begin{eqnarray}
F_{\Omega} & = & 2\left\langle \boldsymbol{B}\right\rangle _{r}\left\langle \boldsymbol{A}\right\rangle _{\phi}\left\langle \boldsymbol{U}\right\rangle _{\phi},\label{eq:EW}\\
F_{U} & = & 2\left\langle \boldsymbol{B}\right\rangle _{r}\left\langle \boldsymbol{A}\right\rangle _{\theta}\left\langle \boldsymbol{U}\right\rangle _{\theta},\label{eq:EU}\\
F_{r}^{\left\langle ab\right\rangle } & = & -\eta_{\chi}\nabla_{r}\left\langle \boldsymbol{a}\cdot\boldsymbol{b}\right\rangle ,\label{eq:fab}\\
F_{\mathcal{\alpha}\beta} & = & 2\left(\alpha_{\phi\phi}+\alpha_{\beta}^{\rm BMR}\right)\left(\left\langle B_{\phi}\right\rangle\left\langle {A}_{\theta}\right\rangle-\left\langle B_{\theta}\right\rangle \left\langle {A}_{\phi}\right\rangle \right)
-2V_{\beta}
\left(\left\langle B_{\theta}\right\rangle \left\langle {A}_{\theta}\right\rangle +\left\langle B_{\phi}\right\rangle \left\langle {A}_{\phi}\right\rangle \right)\label{eq:Fab}\\
F_{\mathcal{\eta}} & = & -2\eta_{T}\left(\left\langle \boldsymbol{A}\right\rangle \times\nabla\times\left\langle \boldsymbol{B}\right\rangle \right)\label{eq:Feta}\\
 & = & -2\eta_{T}\nabla\left(\left\langle \boldsymbol{A}\right\rangle \cdot\left\langle \boldsymbol{B}\right\rangle \right)+2\eta_{T}\left(\left\langle \boldsymbol{A}\right\rangle \cdot\nabla\right)\left\langle {B}_r\right\rangle +2\eta_{T}\left(\left\langle \boldsymbol{B}\right\rangle \cdot\nabla\right)\left\langle {A}\right\rangle_r ,\nonumber 
\end{eqnarray}
where we take into account the isotropic structure of the hydrodynamic
$\alpha$-effect and turbulent diffusion near the solar surface. 
Here, we see that the turbulent diffusion of the dynamo-generated
magnetic field, including the bipolar active regions, produces 
the same type of helicity flux as the diffusive flux of the turbulent
magnetic helicity, $\boldsymbol{F}^{\left\langle ab\right\rangle }=-\eta_{\chi}\boldsymbol{\nabla}\left\langle \boldsymbol{a\cdot}\boldsymbol{b}\right\rangle $.
The part of $F_{\mathcal{\eta}}$ , i.e., $2\eta_{T}\left(\left\langle \boldsymbol{A}\right\rangle \cdot\nabla\right)\left\langle \boldsymbol{B}\right\rangle $
was included in the study of \citet{Hawkes2019AA}. However, the term
$2\eta_{T}\nabla\left(\left\langle \boldsymbol{A}\right\rangle \cdot\left\langle \boldsymbol{B}\right\rangle \right)$
is much larger by magnitude, see Fig.\ref{fig4b}. Also, in our study,
we assume $\hat{\boldsymbol{r}}\cdot\mathrm{d}S=\mathrm{d}\boldsymbol{S}$.
Therefore, the change rate of the magnetic helicity is related to
the helicity change inside the dynamo domain \citep{Berger2000}. 

\section{Boundary conditions}\label{B}

In the solar dynamo models, it is common to employ the vacuum (potential
field) boundary conditions at the top \citet{Krause1980}. Therefore,
in this case, we have $\left\langle  \overline{B}_{\phi}\right\rangle =0$, $\tilde{T}=0$ at the top, the vector-potential  is
$\left\langle \boldsymbol{A}\right\rangle \equiv\left\langle \boldsymbol{A}\right\rangle ^{(p)}$,
where $\left\langle \boldsymbol{A}\right\rangle ^{(p)}$ is the vector-potential
of the potential part of the magnetic field. It satisfies the conditions
$\hat{\mathbf{r}}\cdot\left\langle \boldsymbol{A}\right\rangle ^{(p)}=0$,
and $\nabla\cdot\langle\mathbf{A}\rangle^{(p)}=0$, at the top boundary.
For such boundary conditions, the contribution of the helicity flux that stems from the
twist and tilt magnetic field of BMRs, $F_{\mathcal{\alpha}\beta}\approx0$.
It is noteworthy that in the axisymmetric part of the vector potential
$\left\langle \overline{\boldsymbol{A}}\right\rangle _{\theta}=0$.
Therefore, in the axisymmetric dynamo, the only components of the
 helicity flux are due to the turbulent diffusion, $\overline{F}_{\mathcal{\eta}}=\eta_{T}{\displaystyle \frac{\left\langle \overline{A}_{\phi}\right\rangle }{r}\frac{\partial r\left\langle \overline{B}_{\phi}\right\rangle }{\partial r}}$
and due to the differential rotation, $F_{\Omega}$. Nevertheless,
the impact of this flux on the outer layer is zero because the helicity
of the modeled external magnetic field is zero. The same is true when
we employ a boundary condition for the penetration of the toroidal
magnetic field to the top, e.g., like in the dynamo model of \citet{PK24}.
The consistent study of the helicity flux requires including the coronal
magnetic field and stellar wind in the dynamo simulations, e.g., similar
to simulations of \citet{Warnecke2011,JakBran2021AA,Perri2021ApJ}.

{The less computationally expensive solution can be to consider the
harmonic magnetic field approximation for the outer magnetic field
\citep{Bonanno2016}, i.e., 
\begin{equation}
\left(\nabla^{2}+k^{2}\right)\left\langle \boldsymbol{B}\right\rangle =0,\label{eq:harm}
\end{equation}
for the region $r_{t}<r<2.5R$ and the radial magnetic field for $r\ge2.5R$.
We use $kR=0.1$ as suggested by the results of the above-cited paper.
To connect the external magnetic field with the dynamo region, we employ
the continuity of the normal component of the magnetic field and the
tangential component of the mean electromotive force. For this boundary condition, the
continuity of the tangential component of the mean electromotive force
determines the magnitude of the toroidal magnetic field at the surface.
For the axisymmetric vector potential outside the dynamo domain, we
seek a solution in the form of the decomposition of a product of the spherical
Bessel functions and associated Legendre polynomials as follows (cf
\citealt{Bonanno2016}),
\begin{equation}
A\left(x,\theta,t\right)=\sum A^{(n)}\left(t\right)\frac{\left(\gamma^{(n)}j_{n}\left(x\xi\right)+y_{n}\left(x\xi\right)\right)}{\left(\gamma^{(n)}j_{n}\left(x_{e}\xi\right)+y_{n}\left(x_{e}\xi\right)\right)}\sin\theta P_{n}^{1}\left(\theta\right),\label{eq:Aex}
\end{equation}
where $x=r/R$, $\xi=kR$ and $x_{e}=0.99$ is the external boundary
of the dynamo domain; the constants $A^{(n)}\left(t\right)$ and $\gamma^{(n)}$
are determined by the condition of continuity of the radial magnetic
field at $x_{e}$:
\begin{equation}
\frac{\partial A}{\partial x}=\sum\sin\theta P_{n}^{1}\left(\theta\right)A^{(n)}\left(t\right)\left(\frac{n}{x_{e}}-\xi\frac{\left(\gamma^{(n)}j_{n+1}\left(x_{e}\xi\right)+y_{n+1}\left(x_{e}\xi\right)\right)}{\left(\gamma^{(n)}j_{n}\left(x_{e}\xi\right)+y_{n}\left(x_{e}\xi\right)\right)}\right),\label{eq:Ad}
\end{equation}
and the coronal magnetic field boundary condition, for instance, the
pure radial magnetic field at the radius of the source surface. We
put this point at $x_{s}=2.5$, where we define $\gamma^{(n)}$:
\[
\frac{n}{x_{s}}-\xi\frac{\left(\gamma^{(n)}j_{n+1}\left(x_{s}\xi\right)+y_{n+1}\left(x_{s}\xi\right)\right)}{\left(\gamma^{(n)}j_{n}\left(x_{e}\xi\right)+y_{n}\left(x_{e}\xi\right)\right)}=0.
\]
For the axisymmetric toroidal component, the external magnetic field
decomposition is similar to Eq(\ref{eq:Aex}): 
\begin{equation}
B\left(x,\theta,t\right)=\sum B^{(n)}\left(t\right)\frac{\left(\zeta^{(n)}j_{n}\left(x\xi\right)+y_{n}\left(x\xi\right)\right)}{\left(\zeta^{(n)}j_{n}\left(x_{e}\xi\right)+y_{n}\left(x_{e}\xi\right)\right)}P_{n}^{1}\left(\theta\right),\label{eq:Bex}
\end{equation}
where $\zeta^{(n)}$ is deduced from the condition at $x_{s}$,
$B\left(x_{s},\theta,t\right)=0$. At the top of the dynamo domain,
we require continuity of $\left[\mathcal{E}_{\theta}\right]_{x=x_{e}}=0$
and the same for the toroidal magnetic field. This results in the
following boundary condition 
\begin{equation}
\eta_{T}\frac{\partial xB}{\partial x}=\eta_{T}^{(+)}\sum P_{n}^{1}\left(\theta\right)B^{(n)}\left(t\right)\left((n+1)-\xi x_{e}\frac{\left(\zeta^{(n)}j_{n+1}\left(x_{e}\xi\right)+y_{n+1}\left(x_{e}\xi\right)\right)}{\left(\zeta^{(n)}j_{n}\left(x_{e}\xi\right)+y_{n}\left(x_{e}\xi\right)\right)}\right),\label{eq:Bd}
\end{equation}
where $\eta_{T}^{(+)}$ is the effective turbulent diffusion in the
corona surrounding the dynamo domain. For the case $\eta_{T}^{+}\gg\eta_{T}$
and $\xi,k=0$ , we return to the case of the vacuum boundary conditions.
\citet{Bonanno2016} considered the case $\eta_{T}^{+}=\eta_{T}$
for the advection-dominated dynamo regime with the $\alpha$ effect
concentrated near the bottom of the convection zone. In our model,
the turbulent generation is distributed over the bulk of the convection
zone. For the case $\eta_{T}^{+}=\eta_{T}$, our model shows a strong
surface toroidal magnetic field of about 200 G magnitude. Solar
observations show that in Solar Cycle 24 the surface axisymmetric
toroidal magnetic field was about 1 G \citep{Vidotto2018}. The model
runs in this paper employ the ratio $\eta_{T}^{+}/\eta_{T}=200$, which
results in the magnitude of the surface toroidal magnetic field
about 10 G. Additional studies show that the ratio $\eta_{T}^{+}/\eta_{T}$
affects the dynamo instability threshold for the $\alpha$ effect,
and the increase of $\eta_{T}^{+}/\eta_{T}$ shifts the dynamo threshold
close to the model with the vacuum boundary condition. We hope to
publish the results of that study separately. The boundary conditions
for the superpotentials $\tilde{S}$ and $\tilde{T}$ are considered
in the same way as for the axisymmetric magnetic field components.}

\section{BMR generation functions}.\label{C}

The functions $\xi_{1,2}$ determine the spatial-temporal properties of the emergence of the BMR. They are defined in the same way as \citet{PKT23}: 
\begin{eqnarray}
\!\xi_{1,2}\left(\boldsymbol{r},t,t_{\mathrm{1,2}}\right)\! & = & C_{\beta}\negthinspace\tanh\mathrm{\left(\frac{t}{\tau_{0}}\right)}\exp\left(-m_{\beta}\left(\sin^{2}\!\left(\!\frac{\phi-\phi_{m}}{2}\!\right)\right.\right.\!\label{xib}\\
 &  & \left.\left.+\!\sin^{2}\!\left(\!\frac{\theta-\theta_{\mathrm{m}}}{2}\!\right)\!\right)\right)\psi(r,r_{m},d_{\mathrm{1,2}}),\,0<t<t_{1}\lor t_{1}<t<\delta t\nonumber \\
 & = & 0,\,t>t_{\mathrm{1}}\lor t_{2}>\delta t\nonumber 
\end{eqnarray}
where $\psi$ is a kink-type function of radius, 
\begin{eqnarray}
\psi(r,r_{m},d)\! & =\!\! & \frac{1}{4}\left(\!1\!+\!\mathrm{erf}\left(100\frac{\left(r-r_{m}\right)}{R}\right)\!\right)\label{eq:step}\\
 & \times & \left(\!1\!-\!\mathrm{erf}\left(100\frac{\left(r-(r_{m}+d)\right)}{R}\right)\!\right),
\end{eqnarray}
where $r_{m}$ and $\theta_{m}$ are the radius and the co-latitude
of the BMR's initiations in the convection zone. We set, $t_{1}=\delta t/3$,
where $\delta t=5$ days and the emergence rate $\tau_{0}=1$ day.
For the two-stage process, the emergence time will be $\frac{2}{3}\delta t=4$
days; this corresponds roughly to the emergence parameters of the
large solar active regions \citep{Nortoneal2023}. The parameter $C_{\beta}$
controls the magnitude of the magnetic flux inside BMR; for $C_{\beta}=250$,
we get the simulated BMR's flux of $4\cdot10^{22}$ Mx, when the original
toroidal magnetic field at $r_{m}$ and $\theta_{m}$ is of the strength
1.5\,kG. 
For the source of the initiation of BMR, we take the toroidal magnetic
field in the upper part of the convection zone at the growing stage
of the dynamo cycle, where the condition for the magnetic buoyancy
instability is satisfied \citep{PKT23}.

\bibliographystyle{aasjournal}
\bibliography{dyn}

\end{document}